\documentclass[fleqn,usenatbib]{mnras}
\usepackage{etoolbox}
\makeatletter
\patchcmd\@combinedblfloats{\box\@outputbox}{\unvbox\@outputbox}{}{%
   \errmessage{\noexpand\@combinedblfloats could not be patched}%
}%
 \makeatother
\usepackage{newtxtext,newtxmath}
\usepackage[T1]{fontenc}
\usepackage{ae,aecompl}
\usepackage{graphicx}
\usepackage{amsmath}
\usepackage{amssymb}
\usepackage{times}

\usepackage{hyperref}

\newcommand{\hi}{\ifmmode \ion{H}{I} \else $\ion{H}{I}$ \fi}
\newcommand{\hii}{\ifmmode \ion{H}{II} \else $\ion{H}{II}$ \fi}
\newcommand{\hei}{\ifmmode \ion{He}{I} \else $\ion{He}{I}$ \fi}
\newcommand{\heii}{\ifmmode \ion{He}{II} \else $\ion{He}{II}$ \fi}
\newcommand{\heiii}{\ifmmode \ion{He}{III} \else $\ion{He}{III}$ \fi}
\newcommand{\arepo}{{\sc arepo}\ }
\newcommand{\arepoRT}{{\sc arepo-rt}\ }
\newcommand{\subfind}{{\sc Subfind}\ }
\renewcommand{\vec}[1]{\mathbf{#1}}
\newcommand{\vF}{\vec{F}}
\newcommand{\pp}{\partial}
\newcommand{\pt}{\pp t}

\newcommand{\bbP}{\mathbb{P}}

\usepackage[normalem]{ulem}

\graphicspath{{./}{figures/}}
\usepackage{soul}

\newcommand{\gsim}{\,\lower.7ex\hbox{$\;\stackrel{\textstyle>}{\sim}\;$}}
\newcommand{\lsim}{\,\lower.7ex\hbox{$\;\stackrel{\textstyle<}{\sim}\;$}}

\title[Photoheating feedback due to reionization]{Simulating the effect of photoheating feedback during reionization}
\author[X. H. Wu et al.]{
Xiaohan Wu$^{1}$\thanks{E-mail: xiaohan.wu@cfa.harvard.edu}\href{https://orcid.org/0000-0003-2061-4299}{\includegraphics[scale=0.8]{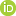}}, 
Rahul Kannan$^{1\thanks{Einstein Fellow}}$,
Federico Marinacci$^{1,2}$\href{https://orcid.org/0000-0003-3816-7028}{\includegraphics[scale=0.8]{figures/orcid.png}},
Mark Vogelsberger$^{3}$
\newauthor
and Lars Hernquist$^{1}$
\\
\\
$^{1}$Harvard-Smithsonian Center for Astrophysics, 60 Garden Street, Cambridge 02138, MA, USA\\
$^{2}$Department of Physics \& Astronomy, University of Bologna, via Gobetti 93/2, 40129 Bologna, Italy\\
$^{3}$Kavli Institute for Astrophysics \& Space Research, Massachusetts Institute of Technology, 77 Massachusetts Ave, Cambridge 02139, \\
MA, USA
}

\date{Accepted 2019 June 12. Received 2019 May 19; in original form 2019 March 13}

\pubyear{2019}

\begin{document}
\label{firstpage}
\pagerange{\pageref{firstpage}--\pageref{lastpage}}
\maketitle

\begin{abstract}
We present self-consistent radiation hydrodynamic simulations of hydrogen reionization performed with \arepoRT complemented by a state-of-the-art galaxy formation model. We examine how photoheating feedback, due to reionization, shapes the galaxies properties. Our fiducial model completes reionization by $z\approx6$ and matches observations of the Ly$\alpha$ forest, the CMB electron scattering optical depth, the high-redshift UV luminosity function, and stellar mass function. Contrary to previous works, photoheating suppresses star formation rates by more than $50\%$ only in halos less massive than $\sim10^{8.4}\ M_\odot$ ($\sim10^{8.8}\ M_\odot$) at $z=6$ $(z=5)$, suggesting inefficient photoheating feedback from photons within galaxies. 
The use of a uniform UV background that heats up the gas at $z\approx10.7$ generates an earlier onset of suppression of star formation compared to our fiducial model. This discrepancy can be mitigated by adopting a UV background model with a more realistic reionization history.
In the absence of stellar feedback, photoheating alone is only able to quench halos less massive than $\sim10^9\ M_\odot$ at $z\gtrsim5$, implying that photoheating feedback is sub-dominant in regulating star formation. In addition, stellar feedback, implemented as a non-local galactic wind scheme in the simulations, weakens the strength of photoheating feedback by reducing the amount of stellar sources. Most importantly, photoheating does not leave observable imprints in the UV luminosity function, stellar mass function, or the cosmic star formation rate density. The feasibility of using these observables to detect imprints of reionization therefore requires further investigation.
\end{abstract}





\section{Introduction}
Cosmological hydrodynamical simulations are among the most powerful tools to study the process of galaxy formation. One of the major challenges for galaxy formation models is to include realistic feedback mechanisms that can regulate gas cooling and star formation. These feedback processes are crucial for simulating realistic galaxy populations \citep[e.g.][]{Voge14a, Scha15}. Among them, stellar feedback in the form of galactic winds driven by supernovae (SNe) is a key ingredient in reducing star formation across a large range of halo masses and is particularly effective in suppressing the faint end slope of the galaxy luminosity function \citep[e.g.][]{Bens03}. Stellar feedback has been invoked in simulations to match the observed galaxy abundances and scaling relations between stars, gas and metals \citep[e.g.][]{Dave06, Dave11a, Dave11b, Voge13}.

Stellar feedback has been extensively studied, producing a comprehensive understanding of its role in galaxy formation. However, at high redshifts ($z\gtrsim5$), photoionization heating due to the reionization process provides another form of feedback, which particularly affects low-mass halos. The epoch of reionization is the era when radiation from the first stars and galaxies reionized the intergalactic medium (IGM), turning it from a cold and neutral medium to a hot and highly ionized one with temperatures of $\sim 20,000 - 30,000$ K \citep[e.g.][]{MER94, McQu12}. During this period, the virial temperatures of halos less massive than $\sim10^9\ M_\odot$ became lower than the mean IGM temperature, leading to suppression of gas accretion onto these objects \citep{Thou96, Gned00, Hoef06, Okam08, Noh14, Katz19}. Moreover, such halos gradually lose their baryon content because their shallow potentials can no longer hold the photoheated gas, leading to an overall reduction of the star formation rate (SFR) of these halos \citep{Petk11, Finl11, Hase13, Ocvi16, Fink19}. Since halos less massive than $\sim10^8\ M_\odot$ can be easily disrupted by a single SN explosion \citep{Finl11}, halos in the mass range $\sim10^8 - 10^9\ M_\odot$ are the most sensitive to the effects of photoheating feedback \citep[for a recent review on the back-reaction of reionization on galaxy formation, see][]{Dayal18}.

Some earlier works studied the mass loss and the suppression of star formation of low-mass halos due to photoheating using a spatially uniform UV background \citep[UVB; e.g.][]{Hoef06, Okam08, Pawl09}. Large-scale cosmological simulations of galaxy formation \citep[e.g.][]{Voge14a, Dubo14, Okam14, Scha15, Dave16, Wein17, Pill18a, Pill18b, Nels18, Spri18, Naim18, Mari18} also often adopt a homogeneous time-varying UVB as an approximation of reionization \citep[e.g.][]{HM12,FG09}. However, reionization is usually believed to be a spatially inhomogeneous and temporally extended process. It is unclear whether a uniform UVB and a patchy reionization produce the same amount of suppression of star formation in low-mass halos. To better address this issue, radiative transfer (RT), or more precisely, radiation hydrodynamics (RHD) simulations, are needed to model the growth of ionized bubbles in a self-consistent manner.

The suppression of star formation in low-mass halos by photoheating feedback has been argued to be observable. For instance, the cosmic star formation rate density (SFRD) may experience a drop during the epoch of reionization \citep{Bark00}. The faint-end slope of the galaxy UV luminosity function (UVLF) may also be sensitive to the reionization history \citep{Gard06}. However, if stellar feedback dominates the regulation of star formation, imprints of photoheating feedback on these observables can become less evident \citep[e.g.][]{Mutch16}. The contribution of low-mass galaxies to reionization may also be reduced \citep[e.g.][]{Wyit13}. A thorough understanding of the interplay between stellar and photoheating feedback requires self-consistent RHD simulations.

A number of previous works have explored the above-mentioned problems using RHD simulations. For instance, \citet{Finl11} and \citet{Ocvi16} both found that at the end of reionization, there is a sharp decrease in SFR of halos less massive than $\sim10^9\ M_\odot$ due to photoheating feedback. In \citet{Ocvi16}, even halos of $10^{10}-10^{11}\ M_\odot$ show a factor of $\sim2$ difference in their SFR when compared to simulations without RT. \citet{Finl11} also illustrated that for low-mass halos ($\lesssim10^9\ M_\odot$), stellar feedback weakens the strength of photoheating feedback on suppressing star formation by reducing the radiation field produced by stellar sources. Contrarily, \citet{Pawl15} found an amplification of the effect of stellar feedback by photoheating feedback \citep[for a larger suite of simulations, see][]{Pawl17}. However, different simulations seem to agree that photoheating is sub-dominant in regulating star formation compared to stellar feedback, and that the latter plays the major role in shaping the galaxy properties \citep[e.g.][]{Rosd18}. The effects of photoheating feedback on observables are also unclear. While \citet{GK14} did not see a drop in the cosmic SFRD or a significant change in the faint end slope of the UVLF in the reionization simulations of \citet{Gned14}, \citet{Finl18} found a small change in the UVLF at ${\rm M_{1500}} > -14$ mag.

In this work we present RHD simulations run with \arepoRT coupled to the Illustris galaxy formation model \citep{Voge13, Voge14a, Voge14b, Genel14, Sija15, Nels15} in order to explore the aforementioned open questions. The Illustris model has been shown to be able to reproduce a number of observed properties of galaxies \citep[e.g.][]{Voge14a, Genel14} and the IGM \citep[e.g.][]{Voge14a, Bird14} at various redshifts, making it a state-of-the-art model for galaxy formation studies. By post-processing the Illustris simulation with RT, \citet{Bauer15} showed that the Illustris star formation history is able to generate a realistic hydrogen reionization history assuming rather low escape fractions ($\lesssim20\%$). In this paper we show how reionization proceeds in the Illustris model when the radiation field is evolved self-consistently with hydrodynamics. We explore how photoheating feedback due to reionization suppresses star formation in halos of different masses. We also analyze the relative importance of stellar feedback and photoheating feedback by comparing simulations with and without stellar feedback. In addition, we assess the differences in the amount of suppression in star formation by performing RHD versus using the uniform \citet{FG09} UVB (hereafter FG09, updated in 2011\footnote{\url{https://galaxies.northwestern.edu/uvb/}}). We briefly examine the feasibility of using the high redshift UVLF and the cosmic SFRD to detect imprints of reionization.

This paper is organized as follows. In Sec.~\ref{sec:methods} we present the galaxy formation model and the RHD scheme. In Sec.~\ref{sec:results} we show the reionization histories in the simulations and discuss the baryon depletion and suppression of star formation in low mass halos due to photoheating feedback. We also explore the implications for observables, including the UVLF, stellar mass function, and cosmic SFRD. We give detailed discussions about the relative importance of photoheating feedback and stellar feedback in Sec.~\ref{sec:discussions}, and summarize our work in Sec.~\ref{sec:conclusions}.

\section{Methods}
\label{sec:methods}
We use the \arepoRT code \citep{Kann18} to solve the coupled equations of gravity, hydrodynamics, and radiative transfer. \arepoRT is a RHD extension of the moving-mesh cosmological hydrodynamic code \arepo \citep{Spri10} which uses an unstructured Voronoi tessellation of the computational domain. The mesh-generating points are allowed to move freely, offering significant flexibility for representing the geometry of the flow. The mesh is then used to solve the equations of ideal hydrodynamics using a second-order unsplit Godunov scheme with an exact Riemann solver. \arepo has been shown to surpass traditional smoothed particle hydrodynamics (SPH) and adaptive mesh refinement (AMR) codes in terms of its accuracy \citep{Keres12, Sija12, Torr12, Voge12, Genel13, Nels13}. Gravitational forces are computed using a Tree-PM scheme \citep{Xu95}, where short-range and long-range forces are calculated using a hierarchical octree algorithm \citep{Barn86} and a Fourier particle-mesh method respectively. We briefly describe the galaxy formation model and the RHD scheme in detail below, which are the key modules of the code needed to perform the simulations presented in this work.

\subsection{Galaxy Formation Model}
\label{sec:GFM}
We adopt the galaxy formation model outlined in \citet{Voge13}. Briefly, gas cells change their internal energy via radiative cooling and heating processes including 
collisional excitation, collisional ionization, recombination, dielectric recombination, bremsstrahlung, Compton cooling off the cosmic microwave background (CMB), photoionization and photoheating \citep{KWH96}. In the original Illustris implementation, gas is assumed to be in ionization equilibrium with the spatially uniform and time-dependent FG09 UVB. The FG09 UVB includes contributions from quasars and star-forming galaxies, the latter dominating at $z\gtrsim3$. It was calibrated to satisfy the observed mean transmission of the Ly$\alpha$ forest at $z=2-4.2$ \citep{FG08a, FG08b}, have \heii reionization by $z\sim3$ \citep{McQu09}, and complete hydrogen reionization by $z=6$. Gas self-shielding is taken into account at $z<6$ by suppressing the photoionization and photoheating rates by a factor of \citep[see][equation A1]{Rahm13}:
\begin{equation}
(1-f)\left[ 1 + \left( \frac{n_{\rm H}}{n_0} \right)^\beta \right]^{\alpha_1} + f\left[ 1 + \frac{n_{\rm H}}{n_0} \right]^{\alpha_2}.
\end{equation}
where $n_{\rm H}$ is the physical hydrogen number density of the cell. The parameters $(\alpha_1, \alpha_2, \beta, f, n_0)$ are linearly interpolated in redshift with the values given in table A1 of \citet{Rahm13}. In the simulations that adopt the FG09 UVB, we continue to use this setup for the treatment of radiative cooling. The RHD implementation of gas cooling adopts a non-equilibrium hydrogen and helium thermochemistry network, which will be presented in Sec.~\ref{sec:RT}.

Gas cooling triggers star formation. We follow the scheme of \citet[][to which we refer the reader for more details]{SH03} and model the star-forming interstellar medium (ISM) gas using an effective equation of state (eEOS). Specifically, we describe the star-forming ISM as a fluid composed of dense cold clouds in pressure equilibrium with an ambient hot gas.
Assuming equilibrium, it can be shown that the effective internal energy per unit mass of the two-phase gas is given by
\begin{equation}
u_{\rm eff} = (1-x) u_{\rm h} + x u_{\rm c},
\label{eq:EOS}
\end{equation}
where $u_{\rm h}$ and $u_{\rm c}$ are the internal energy per unit mass of the hot and cold phases respectively, and $x$ is the mass fraction of the cold gas (computed by the model as a function of gas density). Equation~\ref{eq:EOS} defines the effective equation of state for the star-forming gas.

We consider a gas cell to be star-forming when it exceeds the physical number density threshold\footnote{This is also the density above which the eEOS is imposed.} of $n_{\rm th}\simeq 0.13$ cm$^{-3}$. Following \citet{Spri05}, we determine the temperature of a star-forming gas cell via a weighted mean between the full \citet{SH03} eEOS value and an isothermal EOS at $10^4$ K. In computing the mean, we assign a weight $q = 0.3$ to the eEOS value and, correspondingly, a weight $1 - q$ to the isothermal EOS. For a star-forming gas cell, its star formation timescale is given by
\begin{equation}
t_* = 2.2\sqrt{\frac{n_{\rm th}}{n}}\ {\rm Gyr},
\end{equation}
where $n$ indicates the physical gas number density.
We calculate the SFR as the ratio of the cold gas mass and $t_*$.

We assume that each stellar particle represents a co-eval, single metallicity stellar population which follows a \citet{Chab03} initial mass function (IMF). We calculate mass and metal return due to stellar evolution by integrating over this IMF the time evolution of stellar particles and using information from stellar evolution calculations on the expected main-sequence lifetime, mass return fraction, and heavy element production for a wide range of initial stellar masses and metallicities. We track nine chemical elements -- H, He, C, N, O, Ne, Mg, Si, Fe, and the total gas phase metallicity.

The stellar feedback implementation adopts a non-local energy-driven wind model. In this model winds are directly launched from the star-forming ISM gas in the form of wind particles. After being created, these particles are decoupled from hydrodynamic forces, but not the gravitational forces, until they travel to a region with density below $0.1$ times the density threshold for star formation, or a maximum travel time of $50$ Myr has elapsed. When either of these two criteria is satisfied, we recouple the wind particle and deposit its mass, momentum, thermal energy, and metals into the gas cell where it is currently located. The initial wind velocity $v_{\rm w}$ is
\begin{equation}
v_{\rm w} = \kappa_{\rm w} \sigma^{\rm 1D}_{\rm DM},
\end{equation}
where $\kappa_{\rm w} = 3.7$ is a dimensionless model parameter, and $\sigma^{\rm 1D}_{\rm Dm}$ is the local one-dimensional dark matter velocity dispersion at the current position of the gas cell. We determine the mass carried by galactic winds by computing the mass loading factor $\eta_{\rm w}$, which specifies the ratio of the wind mass flux to the star formation rate:
\begin{equation}
\eta_{\rm w} = \frac{{\rm egy_w}}{v^2_{\rm w}},
\end{equation}
Here ${\rm egy_w} = 1.89\times10^{49}$ erg ${\rm M}^{-1}_\odot$ is the specific energy available for wind generation, i.e. the available Type II SN energy per formed stellar mass. We assume the newly created wind particle has a metallicity that is 0.4 times that of the ambient ISM. The direction of the ejection velocity of wind particles is randomly drawn.

We probabilistically select star-forming gas cells to be converted either into stellar or wind particles, according to the values of star formation rate and wind mass loading factor computed by the model. At each time step $\Delta t$, we draw a random number $x$ from a uniform distribution $U(0,1)$. If $x < 1/(1+\eta_{\rm w})$, we treat the spawning of star particles. Otherwise we consider launching winds. The probability of spawning a star or wind particle of mass $M_*$ from a gas cell of mass $M$ is given by
\begin{equation}
p = \frac{M}{M_*}\left[1 - \exp\left(-\frac{(1+\eta_{\rm w})\Delta t}{t_{\rm SF}}\right)\right]
\end{equation}
where $t_{\rm SF}$ is the ratio of the cell mass and the cell SFR. The star/wind particle mass $M_*$ is set as follows: if $M < 2m_{\rm target}$, then $M_* = M$ and the full gas cell is converted into a star or wind particle. Otherwise the cell only spawns a star (wind) particle of mass $m_{\rm target}$. $m_{\rm target}$ is the mean gas cell mass in the initial conditions (see Table~\ref{tab:sims} for values). We employ a (de-)refinement scheme that keeps the cell masses close (within a factor of 2) to $m_{\rm target}$.

Our simulations are only run until $z=5$, by which redshift hydrogen reionization has completed. Therefore, we do not include metal-line cooling or black hole formation and feedback in our galaxy formation model. This is done for simplicity, but these processes are not expected to have a significant impact at $z\gtrsim5$ \citep[e.g.][]{Ocvi16}. Nevertheless, we use stellar metallicities to calculate the luminosity of star particles, which determines how many photons a star particle should emit and deposit into its surrounding gas cells per unit time (see Sec.~\ref{sec:RT}).

\subsection{Radiative Transfer}
\label{sec:RT}
The RT implementation solves the moment-based radiative transfer equations using the M1 closure relation \citep{Leve84} on a moving mesh \citep[see][for a detailed description of this scheme]{Kann18}. We divide the UV continuum into three frequency bins relevant for hydrogen and helium photoionization: $[13.6, 24.6]$ eV, $[24.6, 54.4]$ eV, $[54.4, 100]$ eV. For each frequency bin $i$, we evolve the comoving photon number density $\tilde{N}_i$ and photon flux $\tilde{\vF}_i$, which are related to the physical photon density $N_i$ and photon flux $F_i$ via
\begin{equation}
\tilde{N}_i = a^3 N_i, \\
\tilde{\vF}_i = a^3 \vF_i,
\end{equation}
where the scale factor $a$ is adopted to account for the loss of photon energy due to cosmological expansion \citep[e.g.][]{Rosd13}. Assuming that the Universe does not expand significantly before a photon is absorbed, the transport equations take the form
\begin{gather}
\frac{\pp \tilde{E}_i}{\pt} + \frac{1}{a}\nabla \cdot \tilde{\vF}_i = 0, \\
\frac{1}{c}\frac{\pp \tilde{\vF}_i}{\pt} + \frac{c}{a}\nabla \cdot \tilde{\bbP}_i = 0,
\end{gather}
where $\tilde{\bbP}_i$ is the radiation pressure tensor and is related to $\tilde{E}_i$ via the Eddington tensor.

We solve photon transport with an explicit scheme, which constrains the simulation time step by the Courant condition. To lower the computational cost we use the reduced speed of light approximation \citep{Gned01} with $\tilde{c} = 0.1c$, where $\tilde{c}$ and $c$ are the reduced and actual speed of light respectively. To further reduce the computing time we perform 32 RT sub-cycles for each hydro time step. In each RT step, we advect radiation by solving the Riemann problem at each cell interface and computing the flux using Godunov's approach \citep{Godu59}. We adopt a Global-Lax-Friedrich flux function \citep{Rusa61}, and achieve second order accuracy by replacing the piecewise constant approximation of Godunov's scheme with a slope limited linear spatial extrapolation and a first order time prediction step to obtain the values of the primitive variables on both sides of the cell interface. We perform the spatial extrapolations using a local least square fit gradient estimate \citep{Pakm16}.

During an RT time step, besides advecting photons, we track the non-equilibrium hydrogen and helium thermochemistry. To do so we use an implicit scheme that takes into account the same radiative processes described in Sec.~\ref{sec:GFM}. We adopt the on-the-spot approximation, assuming that recombination emission is absorbed within the same cell. We trace the ionization fractions of hydrogen and singly and doubly ionized helium for each gas cell. In each hydro time step, we advect these ionization fractions as passive scalars along with the gas.

In our simulations, star particles are the only source of the radiation. We compute the number of photons a star particle emits based on its spectral energy distribution (SED), which is a function of both its age and metallicity, as given by \citet{BC03}. We integrate the SED in each frequency bin to calculate the number of photons to deposit into the surrounding gas cells. Each neighboring gas cell receives a fraction of the total emitted photons of the star particle, proportional to their volume weighted by the evaluation of an SPH cubic spline kernel. The kernel smoothing length is defined as the "standard" SPH smoothing length, which is the length enclosing a predefined number of effective neighbours (in our case $64$). To take into account absorption of photons on unresolved scales, we assume an escape fraction $f_{\rm esc}$ for each star particle, which represents the escape fraction from the birth cloud. We adopt $f_{\rm esc}=0.7$ for all star particles in all the RT simulations. This choice ensures that reionization completes at $z\approx6$ in the simulation with the fiducial stellar feedback model, i.e. the volume-average \hi fraction drops to $\sim10^{-4}$ at $z=6$.

In order to calculate the photoionization and photoheating rates for a gas cell, for each species $j\in[{\rm \hi, \hei, \heii}]$ we compute the mean ionization cross section in each frequency bin $i$ that runs from frequency $\nu_{i,1}$ to $\nu_{i,2}$:
\begin{equation}
\sigma_{ij} = \frac{\displaystyle\int_{\nu_{i,1}}^{\nu_{i,2}} \frac{4\pi J_\nu}{h\nu} \sigma_{\nu j} \mathrm{d}\nu }{\displaystyle\int_{\nu_{i,1}}^{\nu_{i,2}} \frac{4\pi J_\nu}{h\nu} \mathrm{d}\nu },
\end{equation}
where $J_\nu$ is the mean specific intensity. We also calculate the latent heat per photoionization event of species $j$ as:
\begin{equation}
\epsilon_{ij} = \frac{\displaystyle\int_{\nu_{i,1}}^{\nu_{i,2}} \frac{4\pi J_\nu}{h\nu} \sigma_{\nu j} (h\nu - h\nu_{tj}) \mathrm{d}\nu }{ \displaystyle\int_{\nu_{i,1}}^{\nu_{i,2}} \frac{4\pi J_\nu}{h\nu} \sigma_{\nu j} \mathrm{d}\nu },
\end{equation}
where $h\nu_{tj}$ is the ionization potential of the ionic species $j$. In principle $\sigma_{ij}$ and $\epsilon_{ij}$ vary among gas cells due to the different shapes of the spectrum received by the gas cells, which the current code is unable to track. \citet{Rosd13} circumvents this by assuming the same $\sigma_{ij}$ and $\epsilon_{ij}$ for all gas cells and updating them every 10 coarse time steps from the luminosity-weighted averages of the spectra of all star particles in the simulation volume, making $\sigma_{ij}$ and $\epsilon_{ij}$ representative of the average photon population. We note that for the \citet{BC03} spectra, star particles emit most of their photons during the first $\sim5$ Myr of their lifetime, when the calculated $\sigma_{ij}$ and $\epsilon_{ij}$ stay roughly constant and do not change much with metallicity \citep[Fig.~B2 of][]{Rosd13}. We therefore calculate $\sigma_{ij}$ and $\epsilon_{ij}$ using the zero-age zero-metallicity spectrum of the \citet{BC03} model and adopt the same values of $\sigma_{ij}$ and $\epsilon_{ij}$ for all gas cells\footnote{We note that spectral hardening during photon propagation or using a stellar SED model harder than \citet{BC03} can heat the gas to higher temperatures. Appendix~\ref{sec:RSL} will show that a $\sim10,000$ K difference in the halo gas temperature does not have a strong impact on the suppression of halo SFR. Moreover, the IGM temperature is only weakly dependent on the spectral slope \citep{DAlo18Ifront}. We therefore expect harder spectra to have minor effects on our results in Sec.~\ref{sec:results}.}. The resulting $\sigma_{ij}$ and $\epsilon_{ij}$ are tabulated in Table~\ref{tab:group_prop}.

\begin{table*}
\caption{Mean cross sections and photon energies above the ionization thresholds of each species used in the simulations. The energy intervals of the three frequency bins traced in the simulations are indicated in eV in the first column. The other six columns show $\bar{\sigma}_X$ and $\bar{\epsilon}_X$ for each species $X\in[{\rm \hi, \hei, \heii}]$, which are the mean cross section and latent heat per ionization calculated using the zero-age zero-metallicity spectrum of the \protect\citet{BC03} model.}
\label{tab:group_prop}
\begin{tabular}{lcccccc}
\hline
Energy Range & $\bar{\sigma}_{\rm \hi}$ & $\bar{\epsilon}_{\rm \hi}$ & $\bar{\sigma}_{\rm \hei}$ & $\bar{\epsilon}_{\rm \hei}$ & $\bar{\sigma}_{\rm \heii}$ & $\bar{\epsilon}_{\rm \heii}$ \\
$[\rm eV]$ & $[\rm cm^2]$ & $[\rm eV]$ & $[\rm cm^2]$ & $[\rm eV]$ & $[\rm cm^2 ]$ & $[\rm eV]$ \\
\hline
$13.6-24.6$ & $3.2\times10^{-18}$ & 3.4 & 0 & 0 & 0 & 0\\
$23.6-54.4$ & $6.0\times10^{-19}$ & 17.0 & $3.9\times10^{-18}$ & 6.1 & 0 & 0 \\
$54.4-100.0$ & $1.0\times10^{-19}$ & 43.9 & $7.2\times10^{-19}$ & 33.0 & $1.3\times10^{-18}$ & 3.2 \\
\hline
\end{tabular}
\end{table*}

\subsection{Simulations}
\label{subsec:sims}

\begin{table*}
\caption{The simulations used in this work. The table lists the volume side length, choice of stellar feedback, dark matter particle mass, mean gas cell mass in the initial conditions, gravitational softening length, and source of photoheating of each simulation.}
\label{tab:sims}
\begin{tabular}{lcccccc}
\hline
\hline
Name & $L_{\rm box}$ & Winds & $m_{\rm DM}$ & $m_{\rm target}$ & Softening Length & Source of \\
 & $[{\rm cMpc}/h]$ & (Stellar Feedback) & $[M_\odot]$ & $[M_\odot]$ & $[{\rm ckpc}/h]$ & Photoheating \\
\hline
L6n256 fiducial-noRT & 6 & Yes & $1.4\times10^6$ & $2.2\times10^5$ & 0.24 & None \\
L6n256 fiducial-UVB & 6 & Yes & $1.4\times10^6$ & $2.2\times10^5$ & 0.24 & FG09 UVB \\
L6n256 fiducial-RT & 6 & Yes & $1.4\times10^6$ & $2.2\times10^5$ & 0.24 & RT \\
L6n256 NW-noRT & 6 & No & $1.4\times10^6$ & $2.2\times10^5$ & 0.24 & None \\
L6n256 NW-UVB & 6 & No & $1.4\times10^6$ & $2.2\times10^5$ & 0.24 & FG09 UVB \\
L6n256 NW-RT & 6 & No & $1.4\times10^6$ & $2.2\times10^5$ & 0.24 & RT \\
L3n256 fiducial-noRT & 3 & Yes & $1.8\times10^5$ & $2.8\times10^4$ & 0.12 & None \\
L3n256 fiducial-UVB & 3 & Yes & $1.8\times10^5$ & $2.8\times10^4$ & 0.12 & FG09 UVB \\
L3n256 fiducial-RT & 3 & Yes & $1.8\times10^5$ & $2.8\times10^4$ & 0.12 & RT \\
\hline
\end{tabular}
\end{table*}

Table~\ref{tab:sims} summarizes the key features of our simulations. The fiducial simulations have a volume of $(6\ {\rm cMpc}/h)^3$ with $256^3$ dark matter particles and an initial number of $256^3$ gas cells (denoted as L6n256). We run two sets of L6n256 simulations, one using the fiducial stellar feedback model outlined in Sec.~\ref{sec:GFM}, and the other without stellar feedback (tagged as "NW"). For each set of these simulations we run three variations, one without RT or UVB, one with the FG09 UVB, and another with RT, named with postfixes "-noRT", "-UVB", "-RT" respectively. In order to check the convergence of our results, we run three additional simulations with a $3\ {\rm cMpc}/h$ side length and $2\times256^3$ resolution elements (L3n256) using the fiducial stellar feedback model, adopting no RT, FG09 UVB, and RT respectively. Appendix~\ref{sec:convergence} discusses results of the convergence tests.

We adopt a Planck 2016 cosmology with $\Omega_m = 0.3089$, $\Omega_\Lambda = 0.6911$, $\Omega_b = 0.0486$, $h = 0.6774$, and $\sigma_8 = 0.8159$ \citep{Planck16}. Hence the L6n256 and L3n256 simulations have dark matter particle masses of $1.4\times10^6\ M_\odot$ and $1.8\times10^5\ M_\odot$ respectively. The minimum gravitational softening lengths are $0.24$ ckpc$/h$ and $0.12$ ckpc$/h$ in L6n256 and L3n256 respectively. Gas cells use an adaptive softening length tied to the cell radius, limited by a minimum value of $0.03$ ckpc$/h$ in L6n256 and $0.015$ ckpc$/h$ in L3n256 respectively. Our fiducial L6n256 runs are thus able to resolve halos of $10^8\ M_\odot$ with $\sim100$ dark matter particles.

We identify dark matter halos by a friends-of-friends (FOF) algorithm with a minimum particle number of 32 \citep{Davis85} and a linking length of 0.2 times the mean particle separation. Stellar particles and gas cells are attached to these FOF primaries in a secondary linking stage \citep{Dolag09}. We then use the \subfind algorithm to identify gravitationally bound structures \citep{Spri01, Dolag09}. In the following sections, we quote halo mass as the halo virial mass $M_{\rm vir}$, defined as the mass contained in a spherical region with average density that is $200$ times the critical density of the Universe at that time.

\section{Results}
\label{sec:results}

\subsection{Reionization History}

\begin{figure*}
 \includegraphics[width=2\columnwidth]{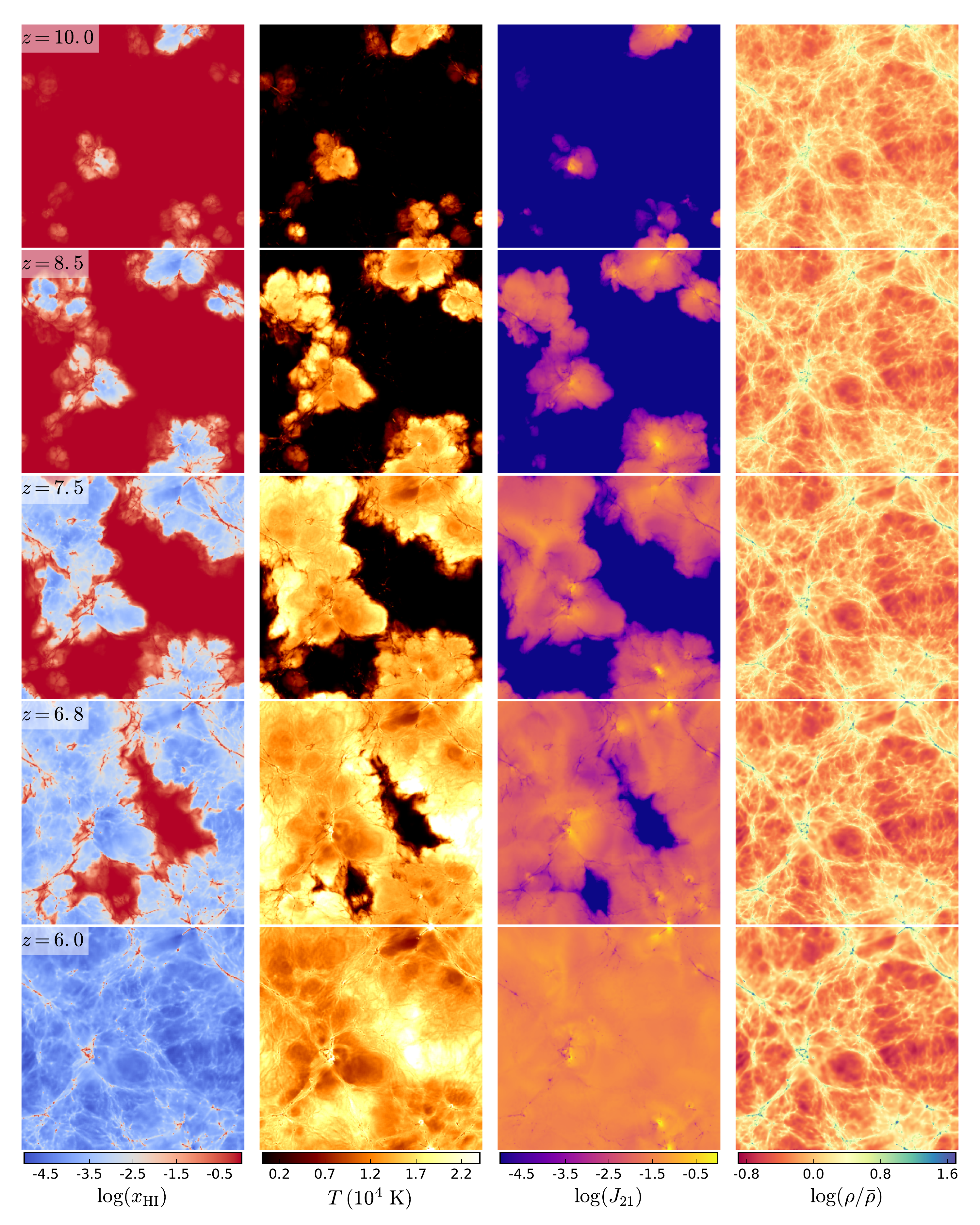}
\caption{A visualization of the reionization process in the L6n256 fiducial-RT run at $z=10.0,8.5,7.5,6.8,6.0$, when the simulated volume is $\sim5\%, \sim20\%, \sim40\%, \sim70\%$, and completely reionized, respectively. The maps have a dimension of $6 \times 6 \times 0.5\ ({\rm cMpc}/h)^3$. From left to right we show the neutral hydrogen fraction, gas temperature, ionizing flux density (in units of $10^{-21}\ {\rm erg\ s^{-1}\ cm^{-2}\ sr^{-1}\ Hz^{-1}}$), and gas density respectively. These maps clearly illustrate how bubbles grow and overlap.}
\label{fig:visualization}
\end{figure*}

We first present how reionization proceeds in our simulations and compare the simulation results with different observational constraints. Fig.~\ref{fig:visualization} shows maps of the \hi fraction, gas temperature, ionizing flux density, and gas density in the fiducial-RT run (from the left to right columns), obtained by projecting a slice of the simulation with a dimension of $6 \times 6 \times 0.5\ ({\rm cMpc}/h)^3$. From top to bottom, the maps are taken at $z=10, 8.5, 7.5, 6.8, 6.0$, respectively.
At $z=10$ the volume-averaged \hi fraction is only about $5\%$. Ionized bubbles form around the early galaxies, which lie on the peaks of the cosmological density field. At $z=8.5$ the global ionized fraction reaches $\sim20\%$, and ionized bubbles are still isolated from each other. By $z=7.5$ the bubbles begin to overlap, and the volume becomes $\sim40\%$ ionized. When the simulated box is $\sim70\%$ ionized at $z=6.8$, only two large neutral islands remain in the slice. Finally complete overlap of the ionized bubbles happens at $z\approx6$. The gas temperature and ionizing flux density evolve in a similar manner, as the ionization fronts sweep through the IGM. Some regions reach temperatures as high as $\sim25000$ K at $z=6.6$. These maps clearly show how the ionized bubbles grow and overlap with each other, illustrating the patchiness of the reionization process.

\begin{figure}
 \includegraphics[width=1\columnwidth]{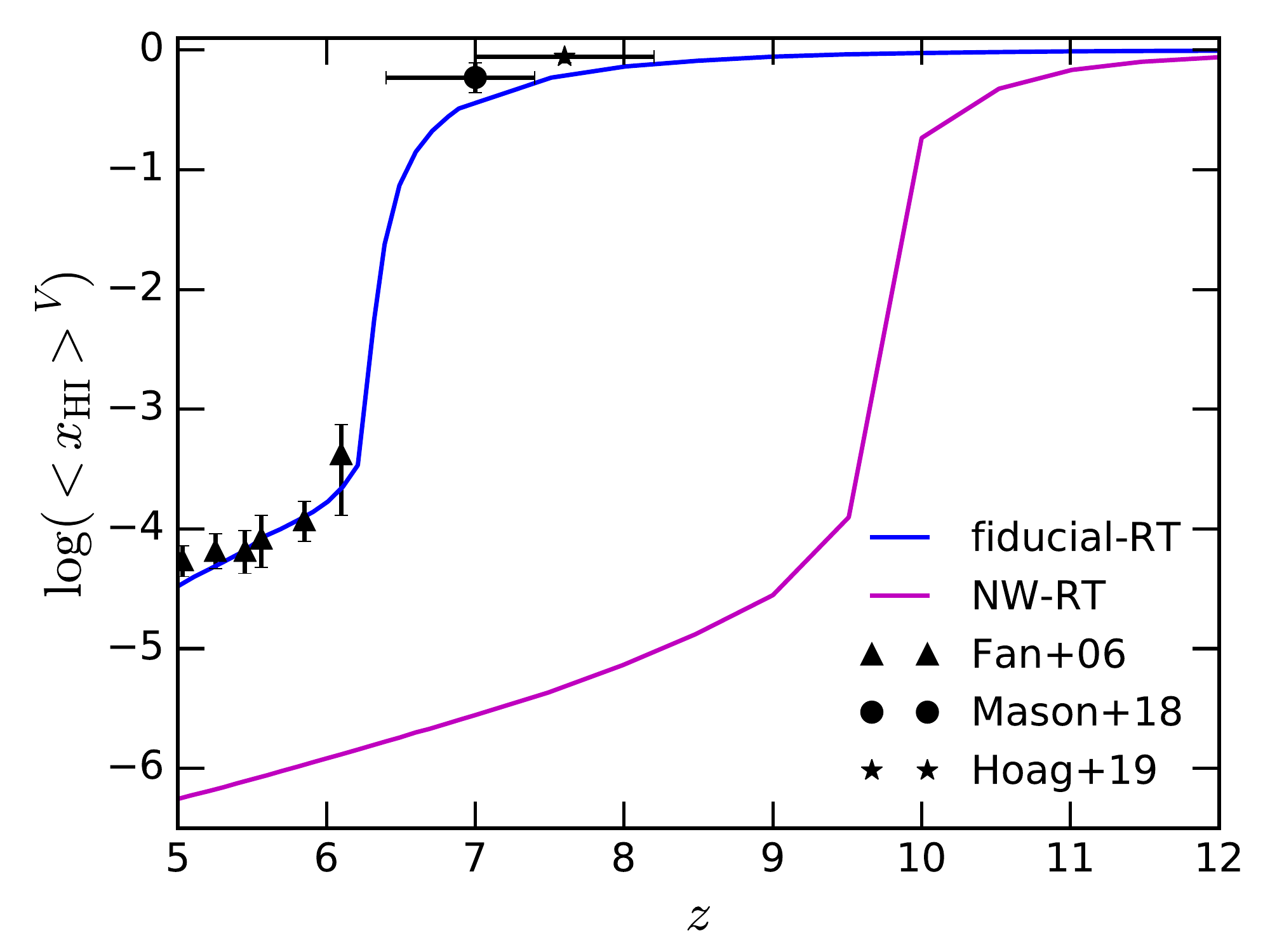}
\caption{Evolution of the volume-averaged neutral hydrogen fraction with redshift in the fiducial-RT (blue line) and NW-RT (magenta line) runs. The fiducial-RT simulation matches the observations of \protect\citet{Fan06} well (black triangles). It also roughly matches the observational data from \protect\citet{Mason18} and \protect\citet{Hoag19} (black circles and stars, respectively).}
\label{fig:HI-z}
\end{figure}

Fig.~\ref{fig:HI-z} illustrates the volume-averaged \hi fraction $\langle x_{\rm \hi} \rangle^V$ as a function of redshift in both the fiducial-RT run (blue line) and the NW-RT run (magenta line), compared with the observations of \citet{Fan06}, \citet{Mason18}, and \citet{Hoag19} (black triangles, circles, and stars, respectively). In the fiducial-RT run, the reionization process is $50\%$ complete at $z\approx7$ and finishes at $z\approx6$, when $\langle x_{\rm \hi} \rangle^V$ drops to $\sim10^{-4}$. $\langle x_{\rm \hi} \rangle^V$ at $z=7-8$ roughly matches the observations, though slightly lower. The post-reionization $\langle x_{\rm \hi} \rangle^V$ matches the observational data well.
We caution, however, that this good match is the consequence of our choices for the values of the escape fraction ($f_{\rm esc} = 0.7$) and reduced speed of light ($\tilde{c} = 0.1 c$). Adopting the actual speed of light with the same $f_{\rm esc}$ would reduce the post-reionization $\langle x_{\rm \hi} \rangle^V$ by a factor of $\sim10$ \citep{Ocvi18RSL, Depa19}.
The NW-RT run has a much earlier-ending reionization and a lower post-reionization $\langle x_{\rm \hi} \rangle^V$. The increased SFR in the NW-RT simulation (about a factor of $10$ higher than the SFR in the fiducial-RT run, see Fig.~\ref{fig:SFR-Mhalo} and Fig.~\ref{fig:sfrdensity}) pushes the time of overlap of ionized bubbles to as early as $z\approx9.5$ and decreases the post-reionization $\langle x_{\rm \hi} \rangle^V$ by about two orders of magnitude.

\begin{figure}
 \includegraphics[width=1\columnwidth]{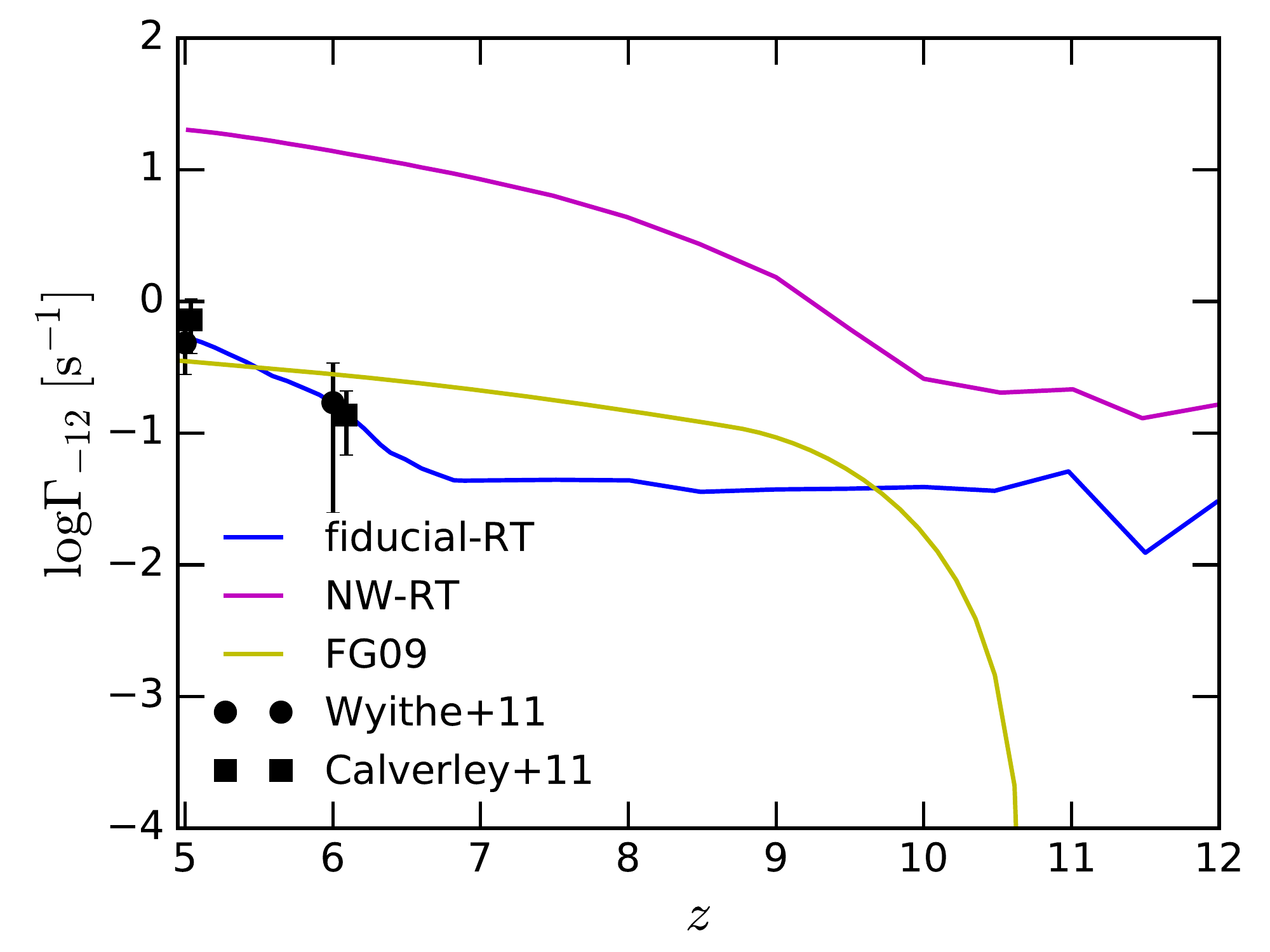}
\caption{Simulated volume-averaged hydrogen photoionization rates in units of $10^{-12}\ {\rm s^{-1}}$ in the fiducial-RT (blue line) and NW-RT (magenta line) simulations. The average is taken using ionized gas cells only (\hi fraction smaller than 50\%). Observational data from \protect\citet[][black circles]{Wyit11} and \protect\citet[][black squares]{Calv11} are shown. The yellow line represents the hydrogen photoionization rate given by the FG09 UVB. The fiducial-RT run roughly matches the observations and the FG09 UVB at $z\sim5-6$.}
\label{fig:UVB}
\end{figure}

Fig.~\ref{fig:UVB} presents the simulated hydrogen photoionization rates ($\Gamma_{\rm \hi}$) compared to the predictions from the FG09 UVB model (yellow line) and the observations of \citet{Wyit11} and \citet{Calv11} (black circles and squares, respectively). $\Gamma_{\rm \hi}$is the volume-weighted average of the hydrogen photoionization rate in ionized gas, defined as any gas cell having a hydrogen ionized fraction larger than $50\%$. The photoionization rate for each gas cell is given by
\begin{equation}
\Gamma_{\rm \hi} = \sum_{i=0}^3 \tilde{c} N_i \sigma_{i, {\rm \hi}}
\end{equation}
where $\tilde{c}$ is the reduced speed of light, $N_i$ is the photon number density of frequency bin $i$, and $\sigma_{i, {\rm \hi}}$ is the hydrogen photoionization cross section of this frequency bin. 
Results from the fiducial-RT simulation match the observational data and the FG09 background at $z\sim5-6$.
However, similar to the behavior of $\langle x_{\rm \hi} \rangle^V$, the use of the actual speed of light with the same $f_{\rm esc}$ would raise the post-reionization UVB amplitude by a factor of $\sim10$ \citep{Ocvi18RSL}.
The NW-RT run, due to the enhanced star formation, has a post-reionization UV background that is nearly two orders of magnitude higher than the fiducial-RT run.

\begin{figure}
 \includegraphics[width=1\columnwidth]{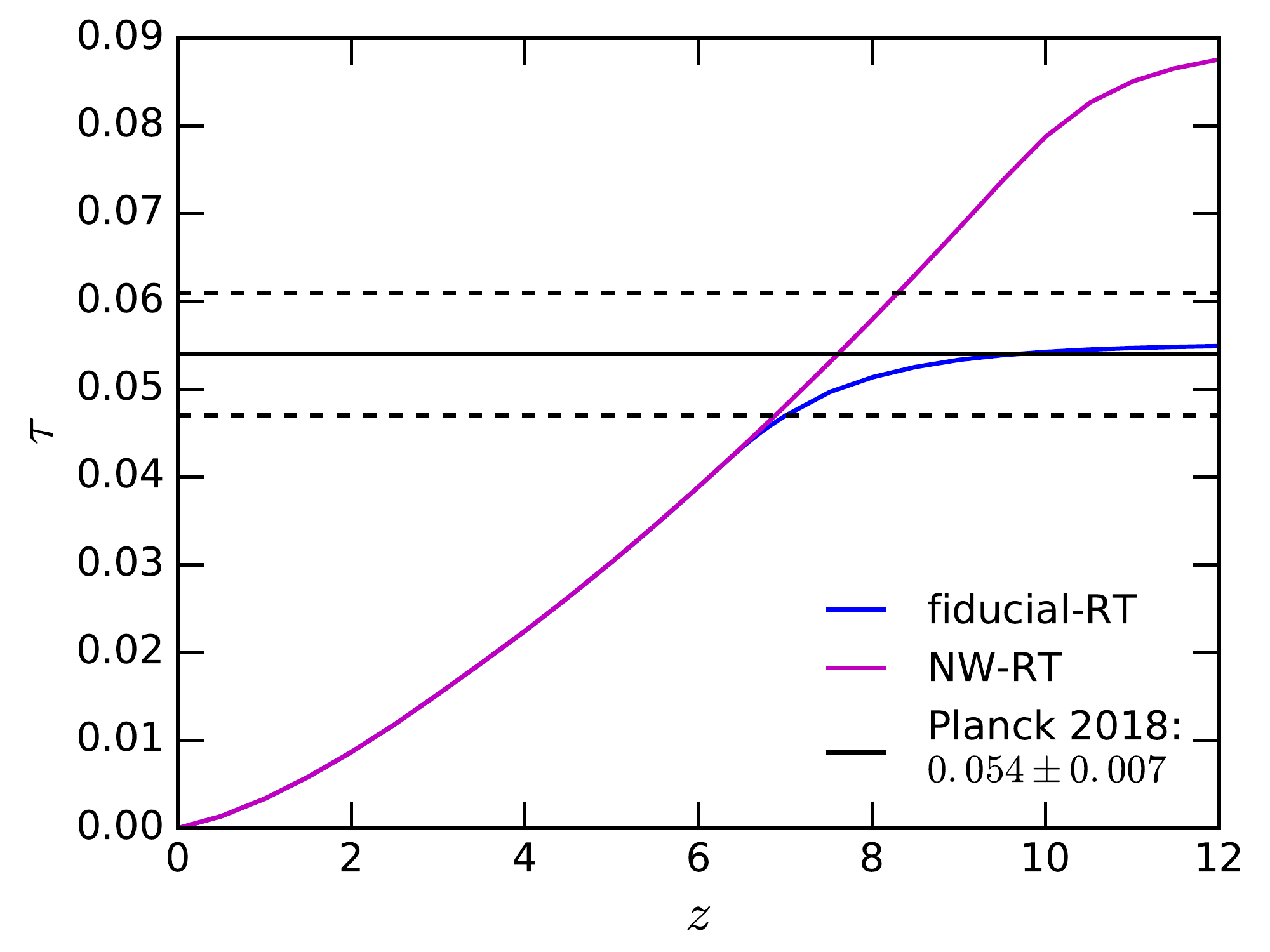}
\caption{Integrated optical depth of Thomson scattering on free electrons as a function of redshift. Blue and magenta lines illustrate results of the fiducial-RT and NW-RT runs, respectively. The observed value and the associated error determined by \protect\citet{Planck18} are shown as the black solid and dashed lines, respectively. Our fiducial model is able to match the \protect\citet{Planck18} results.}
\label{fig:tau}
\end{figure}

The cumulative optical depth to Thomson scattering is another key observable that constrains reionization models. It quantifies the probability of CMB photons scattering off of the free electrons after the epoch of recombination. This optical depth at a redshift $z_0$ is calculated as
\begin{equation}
\tau = c \sigma_{\rm Th} \int_{z_0}^0 n_{\rm e}(z)\frac{\mathrm{d}t}{\mathrm{d}z}\mathrm{d}z,
\end{equation}
where $\sigma_{\rm Th}$ is the Thomson scattering cross section and $n_{\rm e}$ is the number density of free electrons. For our calculations, at $z\ge5$, $n_{\rm e}$ takes the volume-averaged value obtained from the simulations. From $z=5$ to $3$, $n_{\rm e}/n_{\rm H} = 1.08$, since hydrogen reionization is complete and helium is singly ionized. After $z=3$, the time when \heii reionization is usually thought to happen, $n_{\rm e}/n_{\rm H} = 1.158$, assuming full ionization of hydrogen and helium. Fig.~\ref{fig:tau} shows $\tau$ as a function of redshift in the fiducial-RT and NW-RT runs compared to the \citet{Planck18} observations (black lines). A relatively rapid and late-ending reionization in the fiducial-RT run leads to a good match to the observations of \citet{Planck18}. A much earlier and/or a much more extended reionization process increases $\tau$, as seen in the NW-RT run.
More importantly, $\tau$ is relatively insensitive to the choice of the reduced speed of light value because the evolution of $\langle x_{\rm \hi} \rangle^V$ at $\langle x_{\rm \hi} \rangle^V\gtrsim0.01$ is independent of the speed of light \citep[see Appendix~\ref{sec:RSL}, and also][]{Depa19, Ocvi18RSL}.
Our fiducial model therefore is able to roughly match various observational constraints on the hydrogen reionization process.

\subsection{Halo Properties}

\begin{figure*}
\includegraphics[width=2\columnwidth]{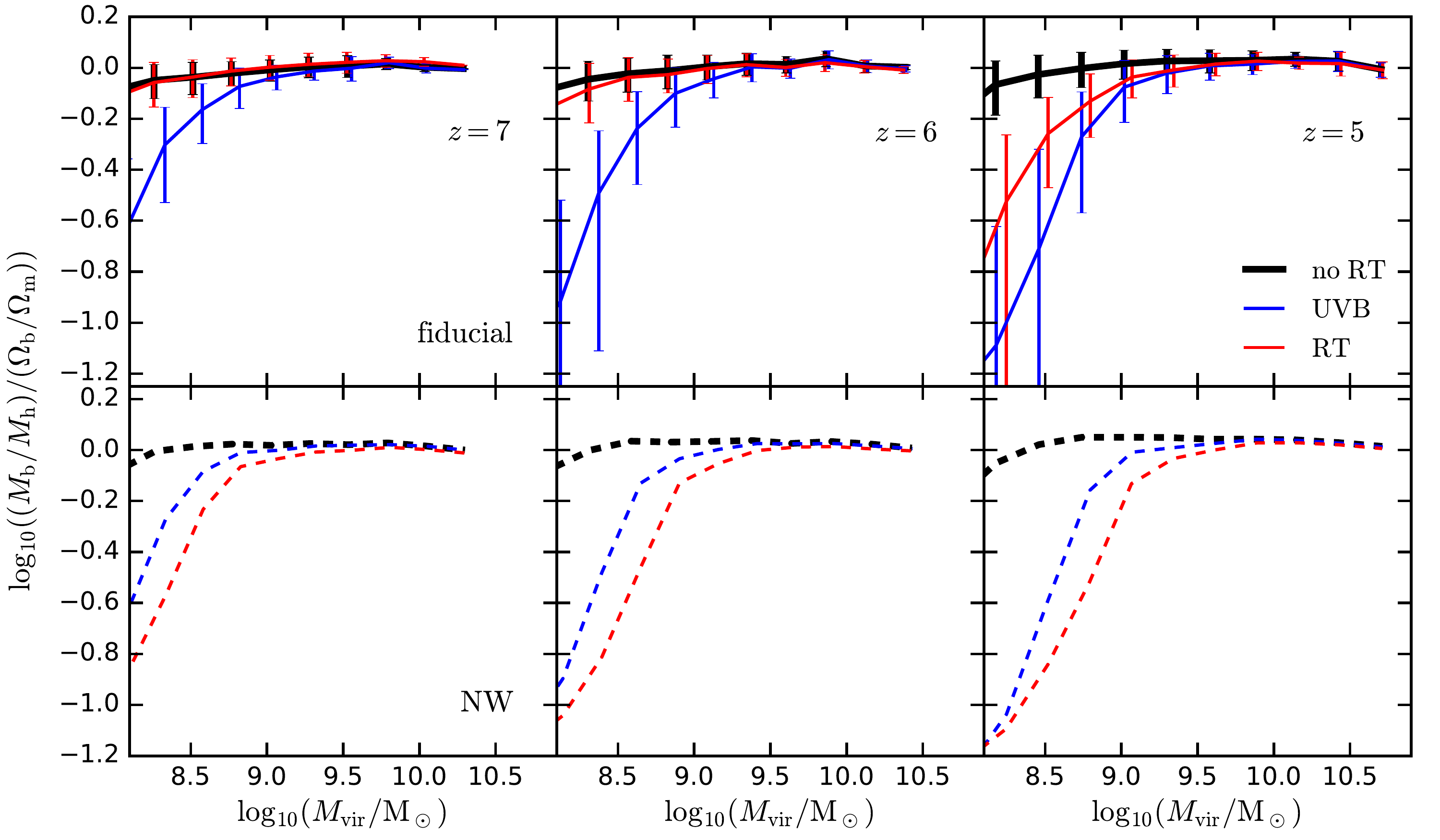}
\caption{Median baryon mass fraction normalized to $\Omega_{\rm b}/\Omega_{\rm m} \sim 0.157$ versus halo mass at $z=7$ (left-hand panels), $6$ (middle panels), and $5$ (right-hand panels). Baryons refer to both gas and stars. The top and bottom panels show results from the fiducial and NW simulations respectively. Lines in black, blue, and red represent simulations without RT nor UVB, with the FG09 UVB, and with RT, respectively. Errorbars represent the $1\sigma$ scatter. To facilitate comparison among the panels we only show errorbars for the fiducial runs. The baryon content is only reduced by less than 0.1 dex in the $10^8\ M_\odot$ halos at $z=6$ in the fiducial-RT run. The NW-RT run is able to deplete the baryons of halos less massive than $10^9\ M_\odot$ at all three redshifts shown. The FG09 UVB suppresses the baryon fraction of $\lesssim10^9\ M_\odot$ halos at all these redshifts, regardless of whether stellar feedback is present.}
\label{fig:fb-Mhalo}
\end{figure*}

We now turn our attention to the effects of photoheating feedback on the properties of low mass ($\lesssim10^9\ M_\odot$) halos. Fig.~\ref{fig:fb-Mhalo} shows the evolution of the median baryon mass fraction as a function of halo mass, at $z=7,6,5$ (in panels from left to right). The baryon fraction is computed as the ratio of the total baryon (gas and stars) mass to the total mass within the halo virial radius, normalized to the global value of $\Omega_{\rm b}/\Omega_{\rm m} \sim 0.157$. Results from the fiducial runs and the NW runs are illustrated in the top and bottom panels, respectively. The black, blue, and red lines show results from the no RT, FG09 UVB, and RT simulations, respectively. Errorbars represent the $1\sigma$ scatter, which are only shown for the fiducial runs to avoid clutter.
In both the fiducial-noRT and NW-noRT simulations, the halos retain roughly all their baryons except the $\sim10^8\ M_\odot$ ones because the fiducial stellar feedback model has outflow velocities that are lower than the escape velocity of the halo \citep{Genel14, Sure15}.
However, both the UVB and RT simulations show a gradual decrease in the baryon fraction of the $\lesssim10^9\ M_\odot$ halos with time due to photoheating. In the fiducial-RT run at $z=7$ when the simulation volume is only $\sim50\%$ ionized, the baryon fraction--halo mass relation is about the same as that of the fiducial-noRT run. Interestingly, at $z=6$ when ionized bubbles have completely overlapped, there is less than $\lesssim0.1$ dex reduction in the baryon content of the $10^8\ M_\odot$ halos in the fiducial-RT run compared to the fiducial-noRT run.  At $z=5$, suppression of the baryon fraction can be seen in halos less massive than $\sim10^9\ M_\odot$, with the $10^8\ M_\odot$ halos having a $\sim80\%$ depletion in their baryons.
This delayed response of the halo gas reservoir to the reionization process indicates that the internal photoheating feedback due to photons in the same halo is not efficient at evaporating gas in our simulations, contrary to the findings of \citet{Hase13}. It also implies that the low-mass halos likely start to be exposed to the ionized bubbles at late stages of reionization, so external photoheating feedback due to photons from other galaxies takes effect late. This external photoheating feedback can also be delayed by gas self-shielding. Indeed, the gas self-shielding threshold at these high redshifts is about 20 times above the cosmic mean \citep{Char18}, indicating that the higher density halo gas is exposed to a less intense UV background than the IGM. The halo gas may therefore require more time to be heated up.

Compared to the fiducial-RT run, there is more suppression of the halo baryon content in the fiducial-UVB run across all redshifts, lowering the baryon fraction of $10^8\ M_\odot$ halos by $75\%$, $88\%$, and $92\%$ at $z=7,6,5$, respectively. After the UVB is turned on at $z\approx10.7$, it quickly heats up all the gas in the simulation volume and thus acts as an early reionization model in terms of its effect on low-mass halos.
Using a UVB model that completes reionization at $z\approx6$ with a more realistic thermal history of the IGM \citep[e.g.][]{Onor17, Puch19} will lead to similar amounts of suppression of the halo baryon content as the fiducial-RT run, which will be demonstrated in Sec.~\ref{sec:discussions}.

Turning off stellar feedback leads to photoheating being able to generate much more suppression of the baryon content of the $M_{\rm vir} \lesssim10^9\ M_\odot$ halos. At $z=7,6,5$, the $10^8\ M_\odot$ halos in the NW-RT run undergo a $85\%,91\%,93\%$ depletion in baryon fraction respectively. On the other hand, the amount of suppression of baryon fraction in the NW-UVB run is the same as in the fiducial-UVB run because the strength of the UVB is independent of the stellar feedback model.
Photoevaporation due to feedback from internal UV photons is thus regulated by the intensity of the UV field. A higher photoionization rate leads to a lower \hi fraction, resulting in less cooling in the dense halo gas because the \hi fraction controls cooling at temperatures of $10^4-10^{5.5}$ K \citep{Ocvi18RSL}. This leads to the possibility that using the actual speed of light may strengthen the suppression of the baryon fraction. However, the photoionization rate is not affected by the reduced speed of light before the overlap of ionized bubbles \citep{Ocvi18RSL}, implying that the choice of the speed of light plays a minor role in determining the suppression of baryon content. We will discuss effects of the reduced speed of light approximation in detail in Sec.~\ref{sec:discussions}. On the other hand, feedback from the external radiation field is less affected by the UVB amplitude, because the IGM thermal evolution is independent of it \citep[e.g.][]{HG97, McQu16}.

\begin{figure*}
\includegraphics[width=2\columnwidth]{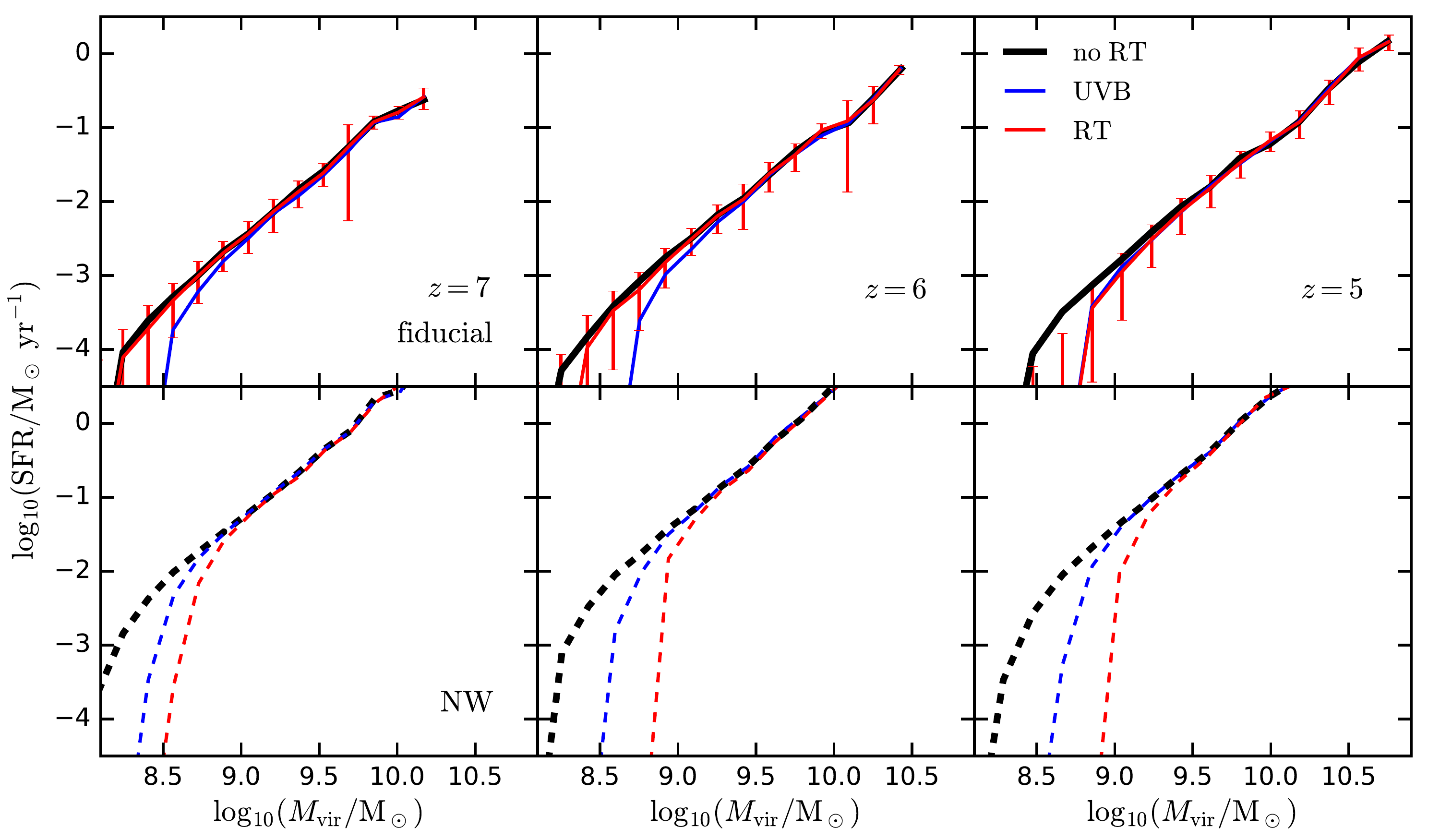}
\caption{Instantaneous star formation rate versus halo mass relations at $z=7, 6, 5$, from the fiducial (top panels) and NW runs (bottom panel). The instantaneous SFR is obtained via summation of the instant SFR of all gas cells inside each halo. The curves indicate median SFR in each halo mass bin. Subplots and line colors are arranged in the same way as in Fig.~\ref{fig:fb-Mhalo}. Errorbars for the fiducial-RT run are plotted to represent $1\sigma$ scatter. At $z=6(5)$ the SFR--halo mass relation of the fiducial-RT run begins deviating from that of the fiducial-noRT run at $\sim10^{8.4}\ M_\odot (\sim10^{8.8}\ M_\odot)$. The FG09 UVB is able to suppress SFR in $\lesssim10^9\ M_\odot$ halos at $z=7-5$. On the other hand, removing stellar feedback leads to quenching in $\lesssim10^9\ M_\odot$ halos in the NW-RT run.}
\label{fig:SFR-Mhalo}
\end{figure*}

The suppression of halo baryon content translates to a similar trend in the suppression of the instantaneous SFR. Fig.~\ref{fig:SFR-Mhalo} presents the median SFR--halo mass relations at $z=7,6,5$ in simulations without RT (black lines), with FG09 UVB (blue lines), and with RT (red lines), performed with both the fiducial (top panels) and NW (bottom panels) models. Errorbars, representing the $1\sigma$ scatter, are only shown for the fiducial-RT run. At $z=7$, no suppression of the halo SFR is seen in the fiducial-RT run. At $z=6$, the SFR--halo mass relation of the fiducial-RT run starts deviating from that of the fiducial-noRT run at about $10^{8.4}\ M_\odot$. In contrast, suppression of SFR at $z=7$ and $6$ is seen in the fiducial-UVB run in halos as massive as $\sim10^9\ M_\odot$. The fiducial-RT run thus only generates a small decrease in the SFR--halo mass relation\footnote{In fact the convergence test in Appendix~\ref{sec:convergence} will show that the L3n256 RT run likely shows no suppression at all at $z=6$.}. At $z=5$ the suppression of SFR in low-mass halos in the fiducial-RT and fiducial-UVB run are comparable. $50\%$ suppression of SFR happens at $M_{\rm vir}\sim10^{8.8}\ M_\odot$ in both simulations, consistent with the findings of previous works \citep[e.g.][]{Finl11, Ocvi16}. However, the lack of suppression of SFR in the fiducial-RT run at the end of reionization ($z=6$) is in tension with these previous studies. 
Since the strength of photoheating feedback relies on the total amount of radiation sources, the extent of reduction in star formation by photoheating feedback is SFR dependent. The no RT simulation in \citet{Finl11} produces an SFR of about $10^{-2.6}\ M_\odot/{\rm yr}$ in $10^{8.5}\ M_\odot$ halos at $z=6$, which is an order of magnitude higher than our fiducial-noRT simulation ($10^{-3.6}\ M_\odot/{\rm yr}$). The higher SFR in their simulations leads to a larger radiation field intensity, allowing photoheating feedback to be more effective at an earlier time. In our simulations stellar feedback is strong enough to suppress star formation efficiently, producing a lower impact of photoheating feedback. Moreover, the recently updated simulations of \citet{Ocvi18} show much less quenching in $10^8-10^9\ M_\odot$ halos compared to the simulations of \citet{Ocvi16} after a recalibration of the star formation sub-grid model, confirming our analysis.

Additionally, both \citet{Ocvi16} and \citet{Finl11} may suffer from insufficient resolution due to large grid sizes. While the $10^8 - 10^9\ M_\odot$ halos have virial radii of $10 - 20$ ckpc$/h$, the gas cells of the \citet{Ocvi16} simulations are of $\sim16$ ckpc$/h$ in width. The finest RT grid in the simulations of \citet{Finl11} is even coarser, having a side length of $\sim93$ ckpc$/h$, so one grid cell could cover an entire low mass halo. A degraded RT grid can smooth out the inhomogeneity of the ionizing background. Thus their RT simulations can possibly mimic the uniform UVB on suppressing halo SFR, making the effect of photoheating feedback seemingly stronger. This shows the importance of resolving these low mass halos both spatially and in mass.

Turning off stellar feedback results in a much larger suppression of SFR in low mass halos due to photoheating feedback, similar to the findings in the baryon fraction--halo mass relation. At $z=7$, the SFR--halo mass relation in the NW-RT run starts deviating from that of the NW-noRT run at $M_{\rm vir} \lesssim 10^{8.7}\ M_\odot$. At $z=6$ and $z=5$ the majority of the $M_{\rm vir} \lesssim10^9\ M_\odot$ halos are completely quenched. The NW-RT run also produces more suppression of SFR than the NW-UVB run at all times because of the increased ionizing radiation intensity. Our findings imply that the strength of photoheating feedback is suppressed by stellar feedback, consistent with \citet{Finl11} but in contrast with \citet{Pawl15}. We will come back to the interplay between stellar feedback and photoheating feedback in Sec.~\ref{sec:discussions} and discuss how this depends on the implementation of the galactic wind.

\subsection{UV Luminosity Function (UVLF), Stellar Mass Function, Cosmic SFR Density}

\begin{figure*}
\includegraphics[width=2\columnwidth]{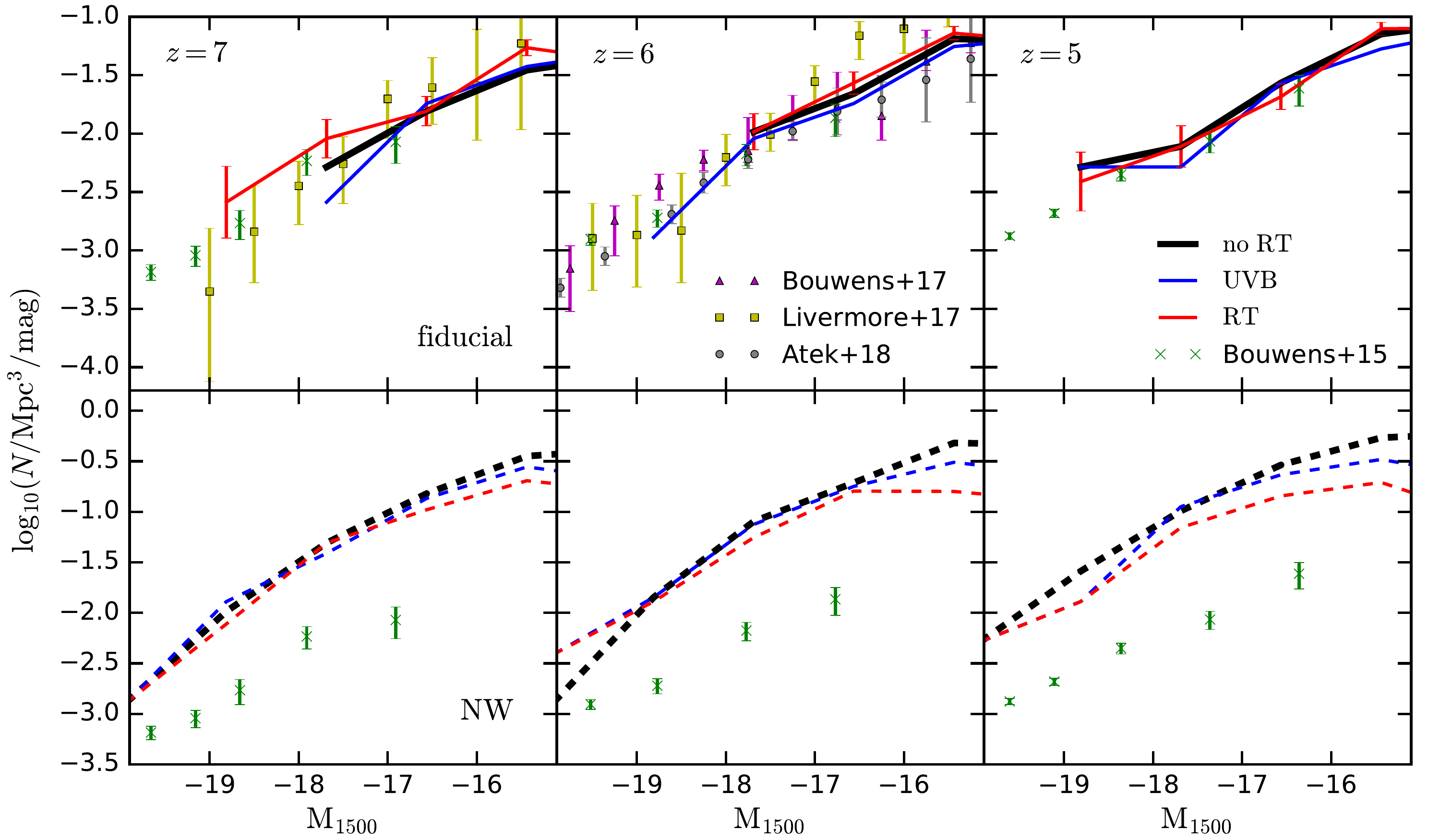}
\caption{Simulated UVLFs at $z=7, 6, 5$ in the fiducial (top panels) and NW (bottom panels) runs. Subplots are arranged in the same way as in Fig.~\ref{fig:fb-Mhalo}. Results from the fiducial model roughly match the observations of \protect\citet{Bouw15} (green crosses), \protect\citet{Bouw17} (magenta triangles), \protect\citet{Live17} (yellow squares), and \protect\citet{Atek18} (gray circles). Error bars are shown for the fiducial-RT run which represent $1\sigma$ scatter. UVLFs in the fiducial simulations are indistinguishable considering the errorbars, regardless of whether RT or UVB is included. Without stellar feedback, the NW-RT and NW-UVB simulations are able to generate a slight flattening of the faint end slope of the UVLFs for magnitudes $\gtrsim-16.5$ at $z=6$ and $5$ compared to the NW-noRT simulation. However, these simulations overproduce the number of galaxies at a given luminosity.}
\label{fig:UVLF}
\end{figure*}

Since the baryon mass fraction--halo mass relation and SFR--halo mass relation are not directly observable, we evaluate how photoheating feedback shapes the more directly observable quantities, including the UVLF, stellar mass function, and cosmic SFR density. The simulated UVLFs are calculated by the following procedure. For each star particle in the \subfind subhalos, we compute its rest-frame spectrum by interpolating the Flexible Stellar Population Synthesis (FSPS) library with nebular emission \citep{Conr09, Conr10} based on its age and metallicity. We do not include dust extinction since it has negligible impact on the UVLF for UV magnitudes $\gtrsim-18$ mag \citep[e.g.][]{Tacc18}. The rest-frame spectrum of each galaxy is then the summation of the spectra of its star particles. The rest-frame 1500 \AA\ luminosity is obtained by convolving the galaxy's spectrum with a top-hat filter centered at 1500 \AA\ with 400 \AA\ in width. 

Fig.~\ref{fig:UVLF} illustrates the simulated UVLFs in the no RT (black lines), FG09 UVB (blue lines), and RT (red lines) runs for both the fiducial (top panels) and NW (bottom panels) models at $z=7,6,5$, compared with the observations of \citet{Bouw15} (green crosses). In the top panels, observational data from \citet{Bouw17} ($z=6$, magenta triangles), \citet{Live17} ($z=7$ and $z=6$, yellow squares), and \citet{Atek18} ($z=6$, gray stars) are also shown. Errorbars, representing the $1\sigma$ scatter, are only plotted for the fiducial-RT run. We cut off the UVLFs at $-15$ mag, because for higher magnitudes the UVLFs will exhibit a turnover caused by lack of resolution (see Appendix~\ref{sec:convergence} for a demonstration). The UVLFs in the fiducial simulations roughly match the observational data at these redshifts, proving the ability of the Illustris galaxy formation model to reproduce high redshift observations. 
There is no observable bend in the faint end slope of the UVLF in the fiducial-RT and fiducial-UVB simulations at ${\rm M_{1500}} < -15$ mag compared to the fiducial-noRT run. Since the most massive galaxy in a $10^9\ M_\odot$ halo in the fiducial simulations is about $10^6\ M_\odot$ in stellar mass, which has a UV magnitude of about $-14$ mag, an observable flattening of the faint end slope of the UVLF is more likely to be seen at ${\rm M_{1500}}\gtrsim-14$ mag.
In principle we can combine the UVLFs from our L3n256 run to push the simulated UVLFs to lower luminosities, but we refrain to do so because the 3 cMpc$/h$ boxes suffer more from cosmic variance. We therefore conclude that the addition of photoheating feedback from reionization does not induce an observable difference in the faint end slope of the $z>5$ UVLF at ${\rm M_{1500}} < -15$ mag.
 
In the NW simulations, the higher level of star formation raises the UV luminosity of a halo of a given mass. Suppression of SFR in the $\lesssim10^9\ M_\odot$ in the NW-RT run is therefore reflected in a suppression of the faint end slope of the UVLF at ${\rm M_{1500}}\gtrsim-16.5$ mag at $z=6$ and $5$ compared to the NW-noRT run, which does not exist in the fiducial runs.

\begin{figure*}
\includegraphics[width=2\columnwidth]{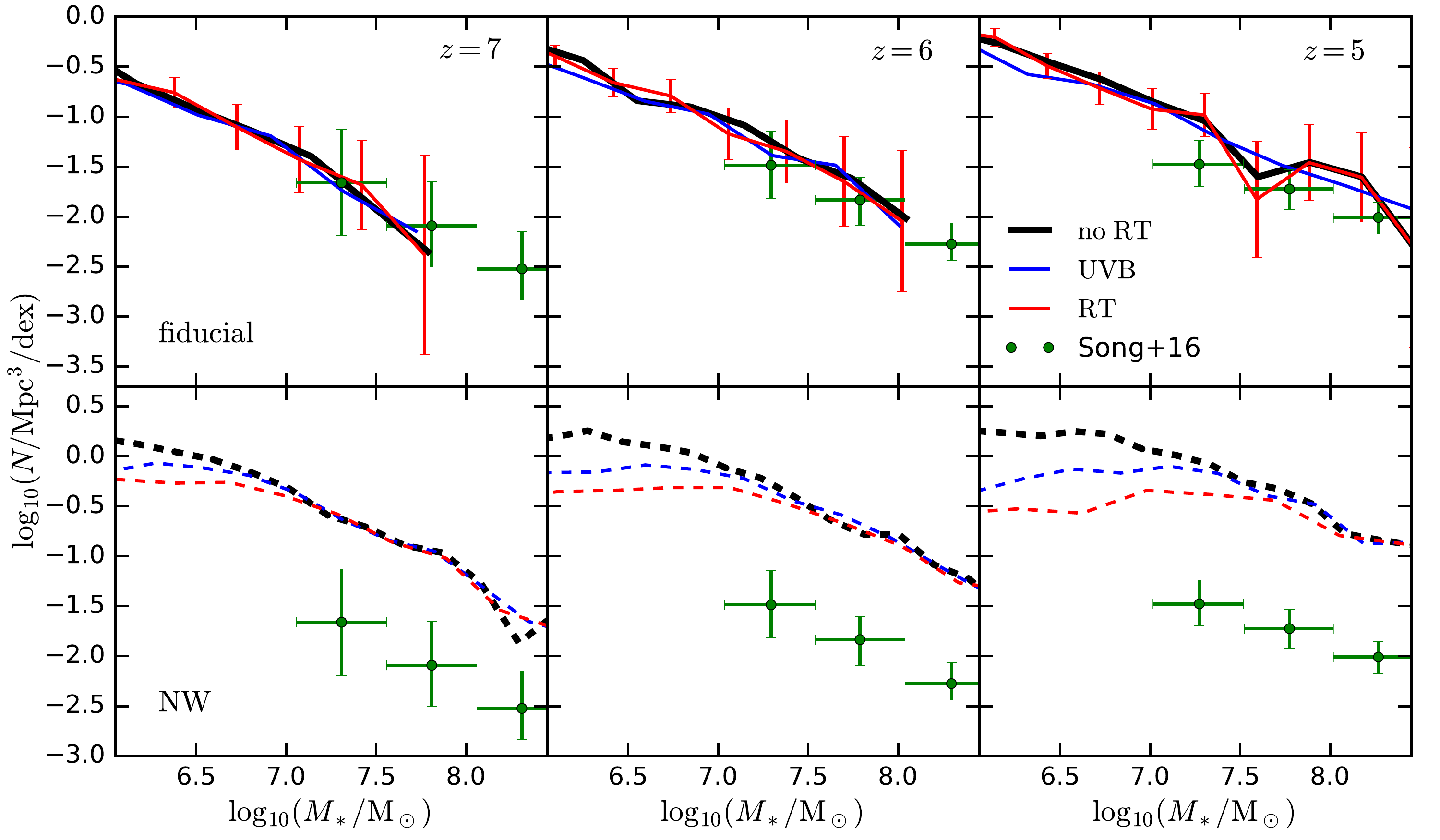}
\caption{Stellar mass functions at $z=7, 6, 5$ in the fiducial (top panels) and NW (lower panels) runs. Subplots are arranged in the same way as in Fig.~\ref{fig:fb-Mhalo}. The fiducial model roughly matches the measurements of \protect\citet[][green crosses]{Song16}. Similar to Fig.~\ref{fig:UVLF}, it is only when stellar feedback is removed that photoheating can generate an observable flattening in the low-mass end (down to $\sim10^6\ M_\odot$) of the stellar mass function.}
\label{fig:stellarMF}
\end{figure*}

Fig.~\ref{fig:stellarMF} presents the simulated stellar mass functions at $z=7,6,5$ compared with the measurements of \citet{Song16} (green circles), which the fiducial runs roughly reproduce\footnote{We note that at high redshifts, the stellar mass function estimate from observations is very uncertain, due to limited sample size and systematic uncertainties in the modeling of galaxy SEDs \citep[see e.g.][for a detailed discussion]{Tacc18}.}. The fiducial-RT and fiducial-UVB runs do not generate any observable suppression of the abundance of low-mass galaxies down to $\sim10^6\ M_\odot$ compared to the fiducial-noRT run. As pointed out earlier, the most massive galaxy in a $10^9\ M_\odot$ halo has about $M_* = 10^6\ M_\odot$ in the fiducial simulations. Hence, the simulated stellar mass function is not expected to show much change at $\gtrsim10^6\ M_\odot$ when photoheating is included. Contrarily, due to the higher stellar mass of galaxies in a given halo in the NW simulations, a suppression of the number of low-mass galaxies is seen at $10^7\ M_\odot$ at $z=7$ in the NW-RT and NW-UVB runs compared to the NW-noRT run, which gets stronger with time. Thus, in the dynamic range that we are able to probe in the simulations, no observable difference is seen in the faint end of the UVLF or the low-mass end of the stellar mass function unless stellar feedback is turned off.

\begin{figure}
\includegraphics[width=1\columnwidth]{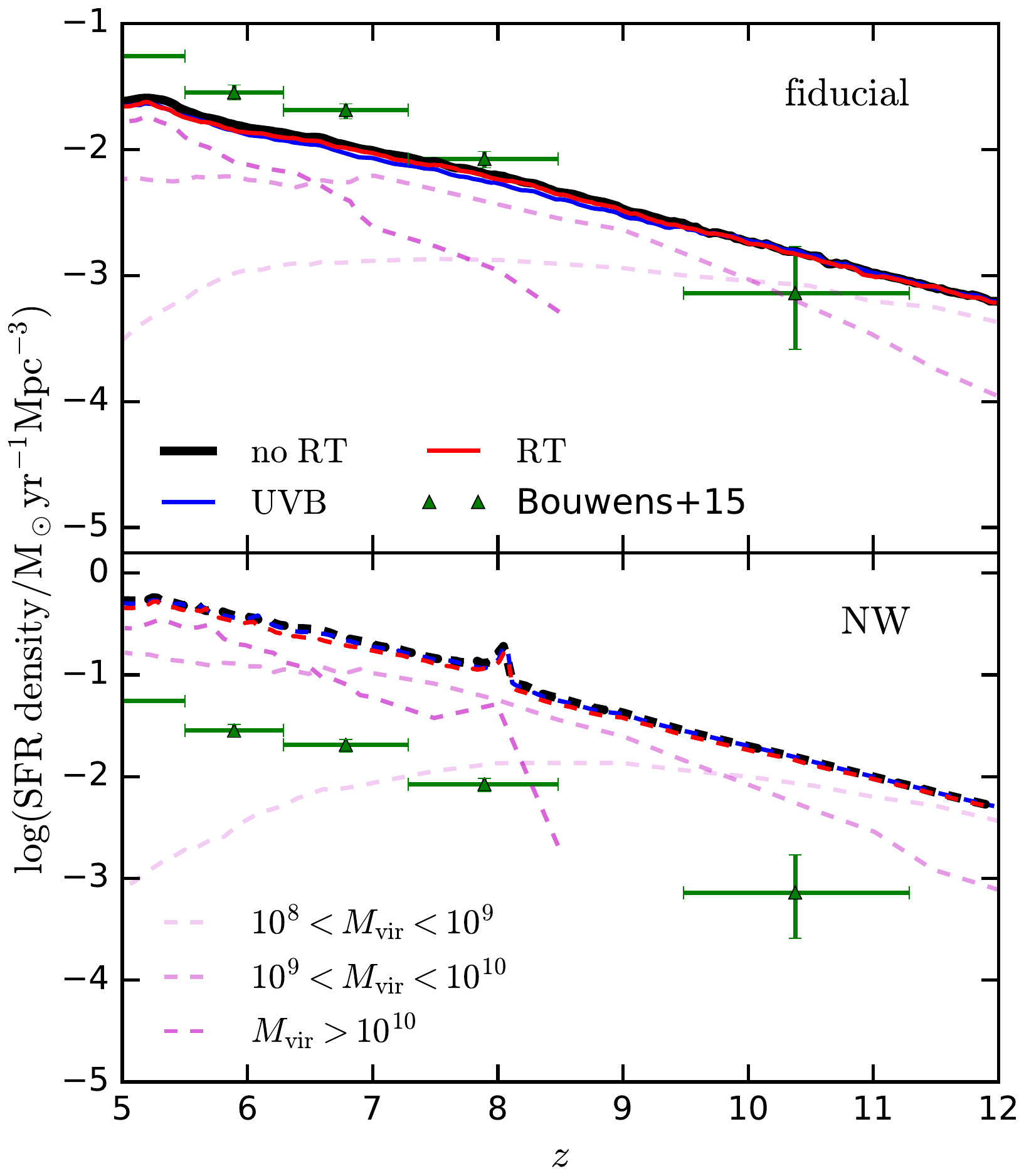}
\caption{The evolution of cosmic SFRD. Top panel shows results from the fiducial runs, which do not match the observations of \protect\citet[][green triangles]{Bouw15} at later times due to the lack of bright sources in the small simulation volume. The bottom panel illustrates results from the NW runs. We show in both panels the star formation histories of three halo mass bins ($10^8 - 10^9\ M_\odot$, $10^9 - 10^{10}\ M_\odot$, $> 10^{10}\ M_\odot$) in the fiducial-RT and NW-RT simulations respectively, using the magenta dashed lines. The sudden increase in the SFR of the $>10^{10}\ M_\odot$ halos causes the bump in the SFRD of the NW runs at $z\approx8$, which is likely stochastic and IC-dependent.} Low-mass halos only dominate the contribution to the cosmic SFRD at $z\gtrsim9.5$, so the suppression of their SFR during reionization does not show up at $z\sim6$ in the cosmic SFRD.
\label{fig:sfrdensity}
\end{figure}

Finally, Fig.~\ref{fig:sfrdensity} illustrates the cosmic SFRD as a function of redshift in the fiducial (top panel) and NW (bottom panel) simulations. The cosmic SFRD in the fiducial runs does not match the observations of \citet{Bouw15} (green triangles), mainly because of the lack of bright sources in the small simulation volume that lowers the cosmic SFRD at late times. Interestingly, there is no observable dip in the cosmic SFRD in either the fiducial-UVB or the fiducial-RT run. Even in the NW-UVB and NW-RT simulations, where the $\lesssim10^9\ M_\odot$ halos are largely quenched due to photoheating feedback, there is no drop in the cosmic SFRD during or after reionization, in contrast with the prediction of \citet{Bark00}. The cosmic SFRD may experience a fall-off if it is dominated by halos of masses $\lesssim10^9\ M_\odot$, which indicates that the reduction of SFR in these halos is not reflected in the cosmic SFRD.
The magenta dashed lines in both panels of Fig.~\ref{fig:sfrdensity} represent the star formation histories of three halo mass bins in the fiducial-RT and NW-RT runs, respectively: $10^8 - 10^9\ M_\odot$, $10^9 - 10^{10}\ M_\odot$, and $> 10^{10}\ M_\odot$. Regardless of whether stellar feedback is included, the $10^8 - 10^9\ M_\odot$ halos dominate the cosmic SFRD before $z\approx9.5$. At $z\sim9.5-6.5$ and $z\lesssim6.5$, the major contribution to the cosmic SFRD comes from the intermediate mass halos and the most massive halos, respectively. Therefore, the dominance of the cosmic SFRD by halos that are not affected by photoheating feedback during and after reionization compensates for the suppressed star formation in the low-mass halos. Our results suggest that it is unlikely that reionization can be probed by an observable dip in the evolution of the cosmic SFRD.

\section{Discussions}
\label{sec:discussions}

\subsection{Interplay Between Photoheating And Stellar Feedback}
Results in Sec.~\ref{sec:results} indicate that stellar feedback is able to suppress the strength of photoheating feedback, because the latter generates much more suppression of halo baryon content and SFR when the former is turned off. In this section we examine in detail the non-linear coupling between the two feedback processes.

\begin{figure*}
\includegraphics[width=2\columnwidth]{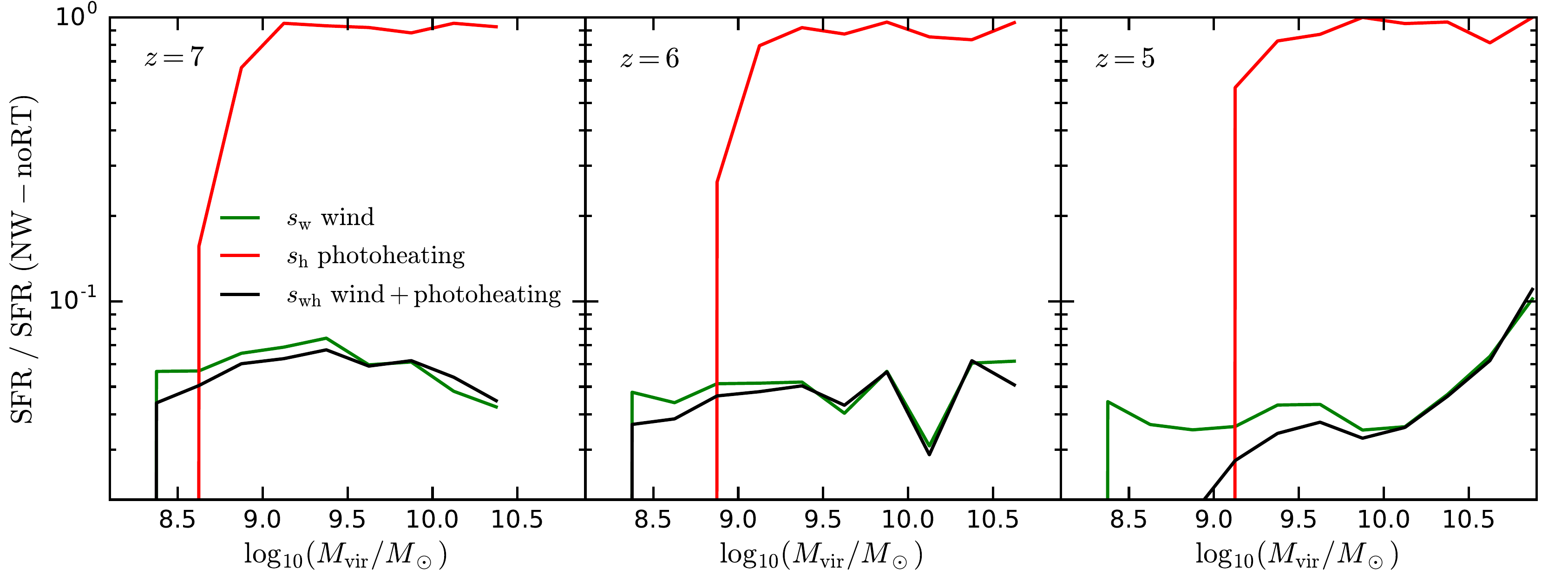}
\caption{Suppression amplitudes of SFR versus halo mass at $z=7,6,5$ due only to photoheating feedback (red lines), stellar feedback (green lines), and caused by the coupled effect of stellar feedback and photoheating feedback (black lines). See the text for a precise definition of these suppression amplitudes. Photoheating only affects the $\lesssim10^9\ M_\odot$ halos and is able to quench them when in the absence of stellar feedback. Its strength is weakened by stellar feedback when the latter dominates the regulation of star formation by reducing the halo SFR by a factor of $\sim20$.}
\label{fig:suppression}
\end{figure*}

To better understand the efficacy of the two feedback mechanisms, we define the amplitude of suppression of SFR due to stellar feedback only as
\begin{equation}
s_{\rm w}(M_{\rm vir}) = \frac{{\rm SFR\ (fiducial\text{-}noRT)}}{{\rm SFR\ (NW\text{-}noRT)}},
\end{equation}
and suppression amplitude caused only by photoheating as
\begin{equation}
s_{\rm h}(M_{\rm vir}) = \frac{{\rm SFR\ (NW\text{-}RT)}}{{\rm SFR\ (NW\text{-}noRT)}}.
\end{equation}
The inclusion of both photoheating and stellar feedback gives a suppression amplitude of
\begin{equation}
s_{\rm wh}(M_{\rm vir}) = \frac{{\rm SFR\ (fiducial\text{-}RT)}}{{\rm SFR\ (NW\text{-}noRT)}}.
\end{equation}

These definitions are the inverse of those in \citet{Pawl09}, but we calculate the suppression amplitudes in this way to avoid division by zero. Fig.~\ref{fig:suppression} shows $s_{\rm w}$ (green), $s_{\rm h}$ (red), $s_{\rm wh}$ (black) as a function of halo mass at $z=7,6,5$ in panels from left to right, respectively. The suppression amplitude $s_{\rm h}$ drops below $\sim0.01$ at halo masses $\lesssim10^{8.4}\ M_\odot$,  $\lesssim10^{8.6}\ M_\odot$, and $\lesssim10^{8.8}\ M_\odot$ at $z=7,6,5$, respectively. For higher mass halos ($\gtrsim10^9\ M_\odot$), the power of photoheating feedback quickly fades away, with the value of $s_{\rm h}$ rapidly increasing to $1$. Photoheating feedback therefore only suppresses star formation in halos of $10^8 - 10^9\ M_\odot$\footnote{We will show in Appendix~\ref{sec:convergence} that this sharp drop in SFR is indeed caused by photoheating feedback, not due to lack of resolution.}. Stellar feedback, in contrast, is able to reduce star formation across the entire halo mass range of $10^8 - 10^{11}\ M_\odot$ by a factor of $\sim20$, making it the dominant mechanism in regulating star formation.

Stellar feedback and photoheating feedback do not seem to amplify the effect of each other when coupled together, contrary to the findings of \citet{Pawl09} and \citet{Pawl15}. While halos less massive than $\sim10^{8.4}\ M_\odot$ and $\sim10^{8.6}\ M_\odot$ are quenched in the NW-RT run at $z=7$ and $6$ respectively, halos in the same mass range in the fiducial-RT run are still forming stars. The value of $s_{\rm wh}$ in these halo mass ranges at $z=7$ and $6$ is at most $\sim0.1$ dex lower than $s_{\rm w}$. If photoheating and stellar feedback boost the power of each other, we should get $s_{\rm wh} < s_{\rm w}s_{\rm h}$, which is not seen. This demonstrates that the strength of photoheating feedback is weakened by stellar feedback, as found in Sec.~\ref{sec:results}. The driving force of this effect is the large difference in the suppression amplitudes $s_{\rm w}$ and $s_{\rm h}$. As stellar feedback dominates the regulation of star formation, it reduces the strength of the radiation field, thus suppressing the impact of photoheating feedback. This effect mainly concerns the internal photoheating feedback, because the external photoheating feedback is less affected by changes in the UVB intensity (see Sec.~\ref{sec:results}). This likely causes the lack of internal photoheating feedback found in Sec.~\ref{sec:results}.

The major reason why we see a different interplay of the two feedback mechanisms from \citet{Pawl09} and \citet{Pawl15} lies in the galactic wind scheme. Ionizing radiation from new born stars heats up the surrounding medium and decreases its density, hence reducing the thermal losses that the wind undergoes after the SNe go off \citep{Stin13, Hopk14, Rosd15, Kann18b}. In local feedback implementations where the SN thermal energy is released into the adjacent gas cells of the star particle, SN feedback works more efficiently when photoheating feedback is included \citep[e.g.][]{Hase13}. By decoupling the wind particles from hydrodynamic forces at $n \gsim 0.01$ cm$^{-3}$, the wind no longer suffers from thermal losses in the high density ISM gas. Effects of photoheating feedback and SN feedback are thus ``decoupled'' in the high density ISM, preventing the boost that photoheating might have on the stength of SN feedback. This also contributes to the insufficient suppression of halo baryon content and SFR at the end of reionization in the fiducial-RT run. Another reason for the disagreement with \citet{Pawl09} is that they used the same UVB \citep{HM01} in simulations with and without stellar feedback, while the UV radiation field can get stronger in the absence of stellar feedback. A larger UVB intensity heats up the dense halo gas more, leading to larger suppression of star formation. We therefore conclude that how stellar feedback and photoheating feedback affect each other is model-dependent.

However, the sub-dominance of photoheating feedback in regulating star formation compared to stellar feedback is in agreement with many other works using different galaxy formation prescriptions. Our findings are consistent with those of \citet{Pawl15} that stellar feedback plays the dominant role in shaping the galaxies properties, and that photoheating does not leave detectable imprints on the UVLF. Moreover, \citet{Rosd18} showed that there is little change in the SFR--halo mass relation when switching from using the single star SED model to binary star SED model. The former failed to complete reionization by $z=6$ in their work, while the latter did. This supports the idea that radiation feedback is sub-dominant in suppressing star formation compared to stellar feedback. Our results are also in agreement with the semi-analytic models of \citet{Wyit13} and \citet{Mutch16}, that stellar feedback plays a greater regulatory role than photoheating. Our predictions on the effects of photoheating on the observables are thus robust. The simulated IGM clumping factor is also relatively robust because the non-local wind scheme does not affect this low-density regime (see Appendix~\ref{sec:clumping}).

\subsection{The Reduced Speed of Light Approximation}

As discussed in Sec.~\ref{sec:results}, using the actual speed of light should boost the post-reionization photoionization rate by a factor of $\sim10$, which possibly leads to more suppression of star formation after reionization. Indeed, Appendix~\ref{sec:RSL} will show that the post-reionization temperatures of gas cells with overdensities of $10-1000$ are lowered by $5,000-10,000$ K when using a reduced speed of light of $0.1c$. By comparing the fiducial-UVB run to a simulation using a scaled version of the FG09 UVB where the photoionization and photoheating rates are raised by a factor of $10$, we found that a $\sim10,000$ K difference in the halo gas temperature results in a $0.1\sim0.2$ dex change in the halo mass threshold where SFR suppression begins to show up. This implies that the post-reionization SFR suppression is not much affected by the adoption of reduced speed of light approximation. On the other hand, the gas temperatures are well-converged when $\langle x_{\rm \hi} \rangle^V$ is above $\sim 0.01$, so the SFR suppression before the overlap of ionized bubbles is also not influenced by the reduced speed of light approximation. Our findings on the SFR suppression are therefore robust to the choice of the reduced speed of light.
\subsection{Other UVB Models}

\begin{figure}
\includegraphics[width=\columnwidth]{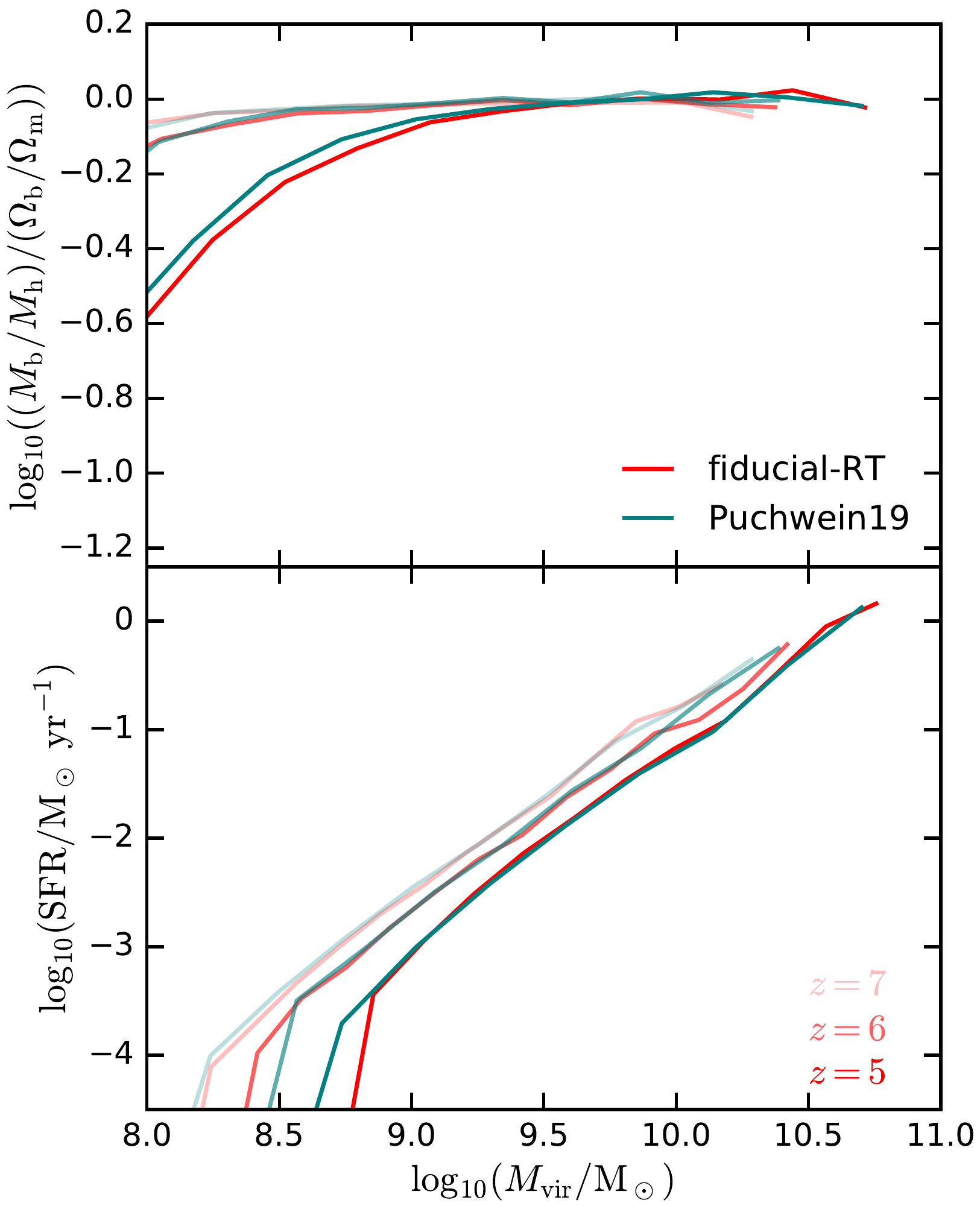}
\caption{Baryon fraction--halo mass relation (top) and SFR--halo mass relation (bottom) at $z=7,6,5$ (colors from light to dark) in the fiducial-RT (red) run and a fiducial simulation using the \protect\citet{Puch19} UVB (teal). The \protect\citet{Puch19} UVB, with a more realistic reionization history, generates similar trends in the suppression of baryon fraction and halo SFR as the fiducial-RT simulation.}
\label{fig:suppression_Puchwein19}
\end{figure}

We explore whether using a UVB model with a more realistic reionization history can generate similar trends in the baryon fraction and SFR suppression as the fiducial-RT simulation. For instance, the \citet{Puch19} UVB model is designed to complete reionization at $z\approx6$ and generates an IGM thermal history with a peak at $z\approx6$. We hence perform an additional fiducial simulation with their ``equivalent-equilibrium'' photoionization and photoheating rates\footnote{\url{https://arxiv.org/src/1801.04931v1/anc/TREECOOL}}. Fig.~\ref{fig:suppression_Puchwein19} presents a comparison of the baryon fraction--halo mass relation (top panel) and SFR--halo mass relation (bottom panel) at $z=7,6,5$ (colors from light to dark), with the fiducial-RT and \citet{Puch19} UVB simulations shown in red and teal, respectively. This UVB model generates similar suppression of the halo baryon fraction and SFR as the fiducial-RT simulation at all redshifts, thus providing comparable external photoheating feedback as the fiducial-RT run.

\section{Conclusions}
\label{sec:conclusions}

In this work we present a suite of state-of-the-art cosmological radiation hydrodynamic simulations with \arepoRT using the Illustris galaxy formation model to simulate the process of reionization. We examined the effects of photoheating feedback due to reionization on galaxy properties and compared the impact of photoheating feedback with that of stellar feedback. Our main results are listed as follows.

\begin{itemize}
\item Reionization completes at $z\approx6$ in the fiducial-RT run. The simulation is able to match the observed volume-averaged neutral hydrogen fraction at $z=5-6$ \citep{Fan06}, the intensity of the post-reionization ionizing background \citep{Calv11, Wyit11}, the cumulative optical depth to Thomson scattering \citep{Planck18}, the UVLFs \citep{Bouw15} and stellar mass functions \citep{Song16} at $z=5-7$, although for the first two there is a dependence on the choice of the reduced speed of light at fixed escape fraction. This demonstrates the ability of our RHD scheme to simulate a realistic reionization process, and the capability of the Illustris galaxy formation model to reproduce high redshift observations.

\item
At $z=6$ ($z=5$), suppression of the baryon content and SFR of low-mass halos ($10^8-10^9\ M_\odot$) due to photoheating feedback only begins to be seen at $\lesssim10^{8.4}\ M_\odot$ ($10^{8.8}\ M_\odot$) in our fiducial-RT run, indicating insufficient internal photoheating feedback from photons in the same halo. However, turning off stellar feedback leads to quenching of these low-mass halos at $z\le7$. The FG09 UVB acts as an early reionization model and begins suppressing star formation earlier. This discrepancy can be mitigated by using a UVB model with a more realistic reionization history \citep[e.g.][]{Puch19}.

\item
Photoheating does not generate any observable flattening in the faint-end slope of the UVLFs up to ${\rm M_{1500}} = -15$ mag in the fiducial simulations, or of the low-mass end of the stellar mass functions down to $10^6\ M_\odot$. However, we point out that there may be an observable difference in the faint-end slope of the UVLF if one can probe down to ${\rm M_{1500}} \gtrsim-14$ mag. We also did not see any dip in the cosmic SFRD during or after reionization, because the SFRD is dominated by halos more massive than $10^{10}\ M_\odot$ which are not affected by photoheating at $z\lesssim6.5$ near the end of reionization.

\item
Photoheating quenches star formation in low-mass halos with masses $\lesssim10^9\ M_\odot$ at $z\gtrsim5$ without the presence of stellar feedback. Its effect on higher mass halos is negligible. On the contrary, stellar feedback is able to reduce star formation across the entire sampled halo mass range by a factor $\sim20$. When coupled together, stellar feedback suppresses the strength of photoheating feedback by reducing the amount of radiation sources. This interplay between the two feedback mechanisms is a result of the non-local galactic wind scheme, but the dominance of stellar feedback in regulating star formation is consistent with other works using different galaxy formation models.
\end{itemize}

In addition to the impact of photoheating being weakened by stellar feedback, another likely cause of the lack of suppression of halo SFR at the end of reionization in the fiducial-RT run is the diversity in the reionization times of halos of different masses. If low-mass halos are exposed to the bulk of the ionized bubbles later than the most massive ones, there may be a delay in the response of their SFR to the reionization process. The evolution of the halo baryon fraction already hints upon this hypothesis (see Sec.~\ref{sec:results}). However, this cannot be checked in our current simulations because we did not include tracer particles that track the reionization time of each gas cell. We defer this analysis and test this scenario in future work.

Future observational facilities, especially the \textit{James Webb Space Telescope} (\textit{JWST}), will be able to observe a number of high redshift galaxies, thus offering new insights into the sources that reionized the Universe. Some of the deepest \textit{JWST} surveys in the first two years can provide a complete sample of galaxies with ${\rm M_{1500}} \lesssim -17$ mag \citep{Will18}, indicating the need for deeper surveys to explore the $z\gtrsim5$ UVLFs at ${\rm M_{1500}} \gtrsim -14$ mag. However, if photoheating only introduces a $\sim0.05$ change in the faint end slope of the UVLF \citep{GK14}, detecting imprints of the reionization process from the faint end slope of the UVLF seems questionable. Moreover, the dominance of $\gtrsim10^{10}\ M_\odot$ halos on the cosmic SFRD at $z\lesssim6.5$ also makes it unlikely to detect an observable dip in the cosmic SFRD during reionization. More careful investigation is therefore needed to explore the feasibility of using these observables to explore reionization.

\section*{Acknowledgements}
We thank Romeel Dav{\'e} and Hui Li for valuable discussions regarding this work, and Sandro Tacchella for generous feedback on the draft of this paper. We also thank Daisuke Nagai, Nick Gnedin, and Kristian Finlator for useful comments on some of the key results of this work. We thank Joop Schaye, Pierre Ocvirk, Stuart Wyithe, Ewald Puchwein, Pratika Dayal, and Steve Finkelstein for important feedback on the first version of this paper. RK acknowledges support from NASA through Einstein Postdoctoral Fellowship grant number PF7-180163 awarded by the Chandra X-ray Center, which is operated by the Smithsonian Astrophysical Observatory for NASA under contract NAS8-03060. FM is supported by the program ``Rita Levi Montalcini'' of the Italian MIUR. MV acknowledges support through an MIT RSC award, a Kavli Research Investment Fund, NASA ATP grant NNX17AG29G, and NSF grants AST-1814053 and AST-1814259.
The simulations were performed on the Harvard computing cluster supported by the Faculty of Arts and Sciences. 




\begin{thebibliography}{99}
\bibitem[Atek et al.(2018)]{Atek18} Atek, H., Richard, J., Kneib, J.-P., \& Schaerer, D.\ 2018, \mnras, 479, 5184 
\bibitem[Bauer et al.(2015)]{Bauer15} Bauer, A., Springel, V., Vogelsberger, M., et al.\ 2015, \mnras, 453, 3593 
\bibitem[Barkana \& Loeb(2000)]{Bark00} Barkana, R., \& Loeb, A.\ 2000, \apj, 539, 20 
\bibitem[Barnes \& Hut(1986)]{Barn86} Barnes, J., \& Hut, P.\ 1986, \nat, 324, 446 
\bibitem[Benson et al.(2003)]{Bens03} Benson, A.~J., Bower, R.~G., Frenk, C.~S., et al.\ 2003, \apj, 599, 38 
\bibitem[Bird et al.(2014)]{Bird14} Bird, S., Vogelsberger, M., Haehnelt, M., et al.\ 2014, \mnras, 445, 2313 
\bibitem[Bouwens et al.(2015)]{Bouw15} Bouwens, R.~J., Illingworth, G.~D., Oesch, P.~A., et al.\ 2015, \apj, 803, 34 
\bibitem[Bouwens et al.(2017)]{Bouw17} Bouwens, R.~J., Oesch, P.~A., Illingworth, G.~D., Ellis, R.~S., \& Stefanon, M.\ 2017, \apj, 843, 129 
\bibitem[Bruzual \& Charlot(2003)]{BC03} Bruzual, G., \& Charlot, S.\ 2003, \mnras, 344, 1000 
\bibitem[Calverley et al.(2011)]{Calv11} Calverley, A.~P., Becker, G.~D., Haehnelt, M.~G., \& Bolton, J.~S.\ 2011, \mnras, 412, 2543 
\bibitem[Chabrier(2003)]{Chab03} Chabrier, G.\ 2003, Publications of the Astronomical Society of the Pacific, 115, 763 
\bibitem[Chardin et al.(2018)]{Char18} Chardin, J., Kulkarni, G., \& Haehnelt, M.~G.\ 2018, \mnras, 478, 1065 
\bibitem[Conroy et al.(2009)]{Conr09} Conroy, C., Gunn, J.~E., \& White, M.\ 2009, \apj, 699, 486 
\bibitem[Conroy \& Gunn(2010)]{Conr10} Conroy, C., \& Gunn, J.~E.\ 2010, \apj, 712, 833 
\bibitem[D'Aloisio et al.(2018)]{DAlo18Ifront} D'Aloisio, A., McQuinn, M., Maupin, O., et al.\ 2018, arXiv:1807.09282 
\bibitem[Dav{\'e} et al.(2006)]{Dave06} Dav{\'e}, R., Finlator, K., \& Oppenheimer, B.~D.\ 2006, \mnras, 370, 273 
\bibitem[Dav{\'e} et al.(2011a)]{Dave11a} Dav{\'e}, R., Oppenheimer, B.~D., \& Finlator, K.\ 2011, \mnras, 415, 11 
\bibitem[Dav{\'e} et al.(2011b)]{Dave11b} Dav{\'e}, R., Finlator, K., \& Oppenheimer, B.~D.\ 2011, \mnras, 416, 1354 
\bibitem[Dav{\'e} et al.(2016)]{Dave16} Dav{\'e}, R., Thompson, R., \& Hopkins, P.~F.\ 2016, \mnras, 462, 3265 
\bibitem[Davis et al.(1985)]{Davis85} Davis, M., Efstathiou, G., Frenk, C.~S., \& White, S.~D.~M.\ 1985, \apj, 292, 371 
\bibitem[Dayal \& Ferrara(2018)]{Dayal18} Dayal, P., \& Ferrara, A.\ 2018, \physrep, 780, 1 
\bibitem[Deparis et al.(2019)]{Depa19} Deparis, N., Aubert, D., Ocvirk, P., et al.\ 2019, \aap, 622, A142.
\bibitem[Dolag et al.(2009)]{Dolag09} Dolag, K., Borgani, S., Murante, G., \& Springel, V.\ 2009, \mnras, 399, 497 
\bibitem[Dubois et al.(2014)]{Dubo14} Dubois, Y., Pichon, C., Welker, C., et al.\ 2014, \mnras, 444, 1453 
\bibitem[Fan et al.(2006)]{Fan06} Fan, X., Strauss, M.~A., Becker, R.~H., et al.\ 2006, \aj, 132, 117 
\bibitem[Faucher-Gigu{\`e}re et al.(2008a)]{FG08a} Faucher-Gigu{\`e}re, C.-A., Prochaska, J.~X., Lidz, A., Hernquist, L., \& Zaldarriaga, M.\ 2008, \apj, 681, 831 
\bibitem[Faucher-Gigu{\`e}re et al.(2008b)]{FG08b} Faucher-Gigu{\`e}re, C.-A., Lidz, A., Hernquist, L., \& Zaldarriaga, M.\ 2008, \apj, 688, 85 
\bibitem[Faucher-Gigu{\`e}re et al.(2009)]{FG09} Faucher-Gigu{\`e}re, C.-A., Lidz, A., Zaldarriaga, M., \& Hernquist, L.\ 2009, \apj, 703, 1416 
\bibitem[Finkelstein et al.(2019)]{Fink19} Finkelstein, S.~L., D'Aloisio, A., Paardekooper, J.-P., et al.\ 2019, arXiv:1902.02792 
\bibitem[Finlator et al.(2011)]{Finl11} Finlator, K., Dav{\'e}, R., \& {\"O}zel, F.\ 2011, \apj, 743, 169 
\bibitem[Finlator et al.(2012)]{Finl12} Finlator, K., Oh, S.~P., {\"O}zel, F., et al.\ 2012, \mnras, 427, 2464.
\bibitem[Finlator et al.(2018)]{Finl18} Finlator, K., Keating, L., Oppenheimer, B.~D., Dav{\'e}, R., \& Zackrisson, E.\ 2018, \mnras, 480, 2628 
\bibitem[Gardner et al.(2006)]{Gard06} Gardner, J.~P., Mather, J.~C., Clampin, M., et al.\ 2006, \ssr, 123, 485 
\bibitem[Genel et al.(2013)]{Genel13} Genel, S., Vogelsberger, M., Nelson, D., et al.\ 2013, \mnras, 435, 1426 
\bibitem[Genel et al.(2014)]{Genel14} Genel, S., Vogelsberger, M., Springel, V., et al.\ 2014, \mnras, 445, 175 
\bibitem[Gnedin(2000)]{Gned00} Gnedin, N.~Y.\ 2000, \apj, 542, 535 
\bibitem[Gnedin \& Abel(2001)]{Gned01} Gnedin, N.~Y., \& Abel, T.\ 2001, \nat, 6, 437 
\bibitem[Gnedin(2014)]{Gned14} Gnedin, N.~Y.\ 2014, \apj, 793, 29 
\bibitem[Gnedin \& Kaurov(2014)]{GK14} Gnedin, N.~Y., \& Kaurov, A.~A.\ 2014, \apj, 793, 30 
\bibitem[Godunov(1959)]{Godu59} Godunov S. K., 1959, Math. Sbornik, 47, 271
\bibitem[Haardt \& Madau(2001)]{HM01} Haardt, F., \& Madau, P.\ 2001, Clusters of Galaxies and the High Redshift Universe Observed in X-rays, 64 
\bibitem[Haardt \& Madau(2012)]{HM12} Haardt, F., \& Madau, P.\ 2012, \apj, 746, 125 
\bibitem[Hasegawa \& Semelin(2013)]{Hase13} Hasegawa, K., \& Semelin, B.\ 2013, \mnras, 428, 154 
\bibitem[Hoag et al.(2019)]{Hoag19} Hoag, A., Brada{\v c}, M., Huang, K.-H., et al.\ 2019, arXiv:1901.09001 
\bibitem[Hoeft et al.(2006)]{Hoef06} Hoeft, M., Yepes, G., Gottl{\"o}ber, S., \& Springel, V.\ 2006, \mnras, 371, 401 
\bibitem[Hopkins et al.(2014)]{Hopk14} Hopkins, P.~F., Kere{\v{s}}, D., O{\~n}orbe, J., et al.\ 2014, \mnras, 445, 581.
\bibitem[Hui \& Gnedin(1997)]{HG97} Hui, L., \& Gnedin, N.~Y.\ 1997, \mnras, 292, 27 
\bibitem[Kannan et al.(2018)]{Kann18b} Kannan, R., Marinacci, F., Simpson, C.~M., et al.\ 2018, arXiv e-prints , arXiv:1812.01614.
\bibitem[Kannan et al.(2019)]{Kann18} Kannan, R., Vogelsberger, M., Marinacci, F., et al.\ 2019, \mnras, 485, 117 
\bibitem[Katz et al.(1996)]{KWH96} Katz, N., Weinberg, D.~H., \& Hernquist, L.\ 1996, \apjs, 105, 19 
\bibitem[Katz et al.(2018)]{Katz18} Katz, H., Kimm, T., Haehnelt, M., et al.\ 2018, \mnras, 478, 4986 
\bibitem[Katz et al.(2019)]{Katz19} Katz, H., Ramsoy, M., Rosdahl, J., et al.\ 2019, arXiv e-prints, arXiv:1905.11414
\bibitem[Kere{\v s} et al.(2012)]{Keres12} Kere{\v s}, D., Vogelsberger, M., Sijacki, D., Springel, V., \& Hernquist, L.\ 2012, \mnras, 425, 2027 
\bibitem[Levermore(1984)]{Leve84} Levermore, C.~D.\ 1984, \jqsrt, 31, 149
\bibitem[Livermore et al.(2017)]{Live17} Livermore, R.~C., Finkelstein, S.~L., \& Lotz, J.~M.\ 2017, \apj, 835, 113 
\bibitem[Marinacci et al.(2018)]{Mari18} Marinacci, F., Vogelsberger, M., Pakmor, R., et al.\ 2018, \mnras, 480, 5113 
\bibitem[Mason et al.(2018)]{Mason18} Mason, C.~A., Treu, T., Dijkstra, M., et al.\ 2018, \apj, 856, 2 
\bibitem[McQuinn et al.(2009)]{McQu09} McQuinn, M., Lidz, A., Zaldarriaga, M., et al.\ 2009, \apj, 694, 842 
\bibitem[McQuinn(2012)]{McQu12} McQuinn, M.\ 2012, \mnras, 426, 1349 
\bibitem[Naiman et al.(2018)]{Naim18} Naiman, J.~P., Pillepich, A., Springel, V., et al.\ 2018, \mnras, 477, 1206 
\bibitem[Nelson et al.(2013)]{Nels13} Nelson, D., Vogelsberger, M., Genel, S., et al.\ 2013, \mnras, 429, 3353 
\bibitem[Nelson et al.(2015)]{Nels15} Nelson, D., Pillepich, A., Genel, S., et al.\ 2015, Astronomy and Computing, 13, 12 
\bibitem[Nelson et al.(2018)]{Nels18} Nelson, D., Pillepich, A., Springel, V., et al.\ 2018, \mnras, 475, 624 
\bibitem[McQuinn \& Upton Sanderbeck(2016)]{McQu16} McQuinn, M., \& Upton Sanderbeck, P.~R.\ 2016, \mnras, 456, 47 
\bibitem[Miralda-Escud{\'e} \& Rees(1994)]{MER94} Miralda-Escud{\'e}, J., \& Rees, M.~J.\ 1994, \mnras, 266, 343 
\bibitem[Mutch et al.(2016)]{Mutch16} Mutch, S.~J., Geil, P.~M., Poole, G.~B., et al.\ 2016, \mnras, 462, 250 
\bibitem[Noh \& McQuinn(2014)]{Noh14} Noh, Y., \& McQuinn, M.\ 2014, \mnras, 444, 503 
\bibitem[Ocvirk et al.(2016)]{Ocvi16} Ocvirk, P., Gillet, N., Shapiro, P.~R., et al.\ 2016, \mnras, 463, 1462 
\bibitem[Ocvirk et al.(2018a)]{Ocvi18RSL} Ocvirk, P., Aubert, D., Chardin, J., et al.\ 2018, arXiv e-prints , arXiv:1803.02434.
\bibitem[Ocvirk et al.(2018b)]{Ocvi18} Ocvirk, P., Aubert, D., Sorce, J.~G., et al.\ 2018, arXiv:1811.11192 
\bibitem[Okamoto et al.(2008)]{Okam08} Okamoto, T., Gao, L., \& Theuns, T.\ 2008, \mnras, 390, 920 
\bibitem[Okamoto et al.(2014)]{Okam14} Okamoto, T., Shimizu, I., \& Yoshida, N.\ 2014, \pasj, 66, 70 
\bibitem[O{\~n}orbe et al.(2017)]{Onor17} O{\~n}orbe, J., Hennawi, J.~F., \& Luki{\'c}, Z.\ 2017, \apj, 837, 106.
\bibitem[Pakmor et al.(2016)]{Pakm16} Pakmor, R., Springel, V., Bauer, A., et al.\ 2016, \mnras, 455, 1134 
\bibitem[Pawlik \& Schaye(2009)]{Pawl09} Pawlik, A.~H., \& Schaye, J.\ 2009, \mnras, 396, L46 
\bibitem[Pawlik et al.(2015)]{Pawl15} Pawlik, A.~H., Schaye, J., \& Dalla Vecchia, C.\ 2015, \mnras, 451, 1586 
\bibitem[Pawlik et al.(2017)]{Pawl17} Pawlik, A.~H., Rahmati, A., Schaye, J., Jeon, M., \& Dalla Vecchia, C.\ 2017, \mnras, 466, 960 
\bibitem[Petkova \& Springel(2011)]{Petk11} Petkova, M., \& Springel, V.\ 2011, \mnras, 412, 935 
\bibitem[Pillepich et al.(2018a)]{Pill18a} Pillepich, A., Springel, V., Nelson, D., et al.\ 2018, \mnras, 473, 4077 
\bibitem[Pillepich et al.(2018b)]{Pill18b} Pillepich, A., Nelson, D., Hernquist, L., et al.\ 2018, \mnras, 475, 648 
\bibitem[Planck Collaboration et al.(2016)]{Planck16} Planck Collaboration, Ade, P.~A.~R., Aghanim, N., et al.\ 2016, \aap, 594, A13
\bibitem[Planck Collaboration et al.(2018)]{Planck18} Planck Collaboration, Aghanim, N., Akrami, Y., et al.\ 2018, arXiv:1807.06209 
\bibitem[Puchwein et al.(2019)]{Puch19} Puchwein, E., Haardt, F., Haehnelt, M.~G., et al.\ 2019, \mnras, 485, 47.
\bibitem[Rahmati et al.(2013)]{Rahm13} Rahmati, A., Pawlik, A.~H., Rai{\v c}evi{\'c}, M., \& Schaye, J.\ 2013, \mnras, 430, 2427 
\bibitem[Rosdahl et al.(2013)]{Rosd13} Rosdahl, J., Blaizot, J., Aubert, D., Stranex, T., \& Teyssier, R.\ 2013, \mnras, 436, 2188 
\bibitem[Rosdahl et al.(2015)]{Rosd15} Rosdahl, J., Schaye, J., Teyssier, R., et al.\ 2015, \mnras, 451, 34.
\bibitem[Rosdahl et al.(2018)]{Rosd18} Rosdahl, J., Katz, H., Blaizot, J., et al.\ 2018, \mnras, 479, 994 
\bibitem[Rusanov(1961)]{Rusa61} Rusanov V. V., 1961, J. Comput. Math. Phys. USSR, 1, 267
\bibitem[Schaye et al.(2015)]{Scha15} Schaye, J., Crain, R.~A., Bower, R.~G., et al.\ 2015, \mnras, 446, 521 
\bibitem[Sijacki et al.(2012)]{Sija12} Sijacki, D., Vogelsberger, M., Kere{\v s}, D., Springel, V., \& Hernquist, L.\ 2012, \mnras, 424, 2999 
\bibitem[Sijacki et al.(2015)]{Sija15} Sijacki, D., Vogelsberger, M., Genel, S., et al.\ 2015, \mnras, 452, 575 
\bibitem[Song et al.(2016)]{Song16} Song, M., Finkelstein, S.~L., Ashby, M.~L.~N., et al.\ 2016, \apj, 825, 5 
\bibitem[Springel et al.(2001)]{Spri01} Springel, V., White, S.~D.~M., Tormen, G., \& Kauffmann, G.\ 2001, \mnras, 328, 726 
\bibitem[Springel \& Hernquist(2003)]{SH03} Springel, V., \& Hernquist, L.\ 2003, \mnras, 339, 289 
\bibitem[Springel et al.(2005)]{Spri05} Springel, V., Di Matteo, T., \& Hernquist, L.\ 2005, \mnras, 361, 776 
\bibitem[Springel(2010)]{Spri10} Springel, V.\ 2010, \mnras, 401, 791 
\bibitem[Springel et al.(2018)]{Spri18} Springel, V., Pakmor, R., Pillepich, A., et al.\ 2018, \mnras, 475, 676 
\bibitem[Stinson et al.(2013)]{Stin13} Stinson, G.~S., Brook, C., Macci{\`o}, A.~V., et al.\ 2013, \mnras, 428, 129.
\bibitem[Suresh et al.(2015)]{Sure15} Suresh, J., Bird, S., Vogelsberger, M., et al.\ 2015, \mnras, 448, 895 
\bibitem[Tacchella et al.(2018)]{Tacc18} Tacchella, S., Bose, S., Conroy, C., Eisenstein, D.~J., \& Johnson, B.~D.\ 2018, \apj, 868, 92 
\bibitem[Torrey et al.(2012)]{Torr12} Torrey, P., Vogelsberger, M., Sijacki, D., Springel, V., \& Hernquist, L.\ 2012, \mnras, 427, 2224 
\bibitem[Thoul \& Weinberg(1996)]{Thou96} Thoul, A.~A., \& Weinberg, D.~H.\ 1996, \apj, 465, 608 
\bibitem[Vogelsberger et al.(2012)]{Voge12} Vogelsberger, M., Sijacki, D., Kere{\v s}, D., Springel, V., \& Hernquist, L.\ 2012, \mnras, 425, 3024 
\bibitem[Vogelsberger et al.(2013)]{Voge13} Vogelsberger, M., Genel, S., Sijacki, D., et al.\ 2013, \mnras, 436, 3031 
\bibitem[Vogelsberger et al.(2014a)]{Voge14a} Vogelsberger, M., Genel, S., Springel, V., et al.\ 2014, \nat, 509, 177 
\bibitem[Vogelsberger et al.(2014b)]{Voge14b} Vogelsberger, M., Genel, S., Springel, V., et al.\ 2014, \mnras, 444, 1518 
\bibitem[Weinberger et al.(2017)]{Wein17} Weinberger, R., Springel, V., Hernquist, L., et al.\ 2017, \mnras, 465, 3291 
\bibitem[Williams et al.(2018)]{Will18} Williams, C.~C., Curtis-Lake, E., Hainline, K.~N., et al.\ 2018, \apjs, 236, 33 
\bibitem[Wyithe \& Bolton(2011)]{Wyit11} Wyithe, J.~S.~B., \& Bolton, J.~S.\ 2011, \mnras, 412, 1926 
\bibitem[Wyithe \& Loeb(2013)]{Wyit13} Wyithe, J.~S.~B., \& Loeb, A.\ 2013, \mnras, 428, 2741 
\bibitem[Xu(1995)]{Xu95} Xu, G.\ 1995, \apjs, 98, 355 

\end{thebibliography}

\bibliographystyle{mnras}




\appendix

\section{Numerical convergence of the halo SFR and UV luminosity function}
\label{sec:convergence}
In this section we discuss the numerical convergence of the of our simulations by using the L6n256 L3n256 boxes, both run with the fiducial model.

\begin{figure}
\includegraphics[width=\columnwidth]{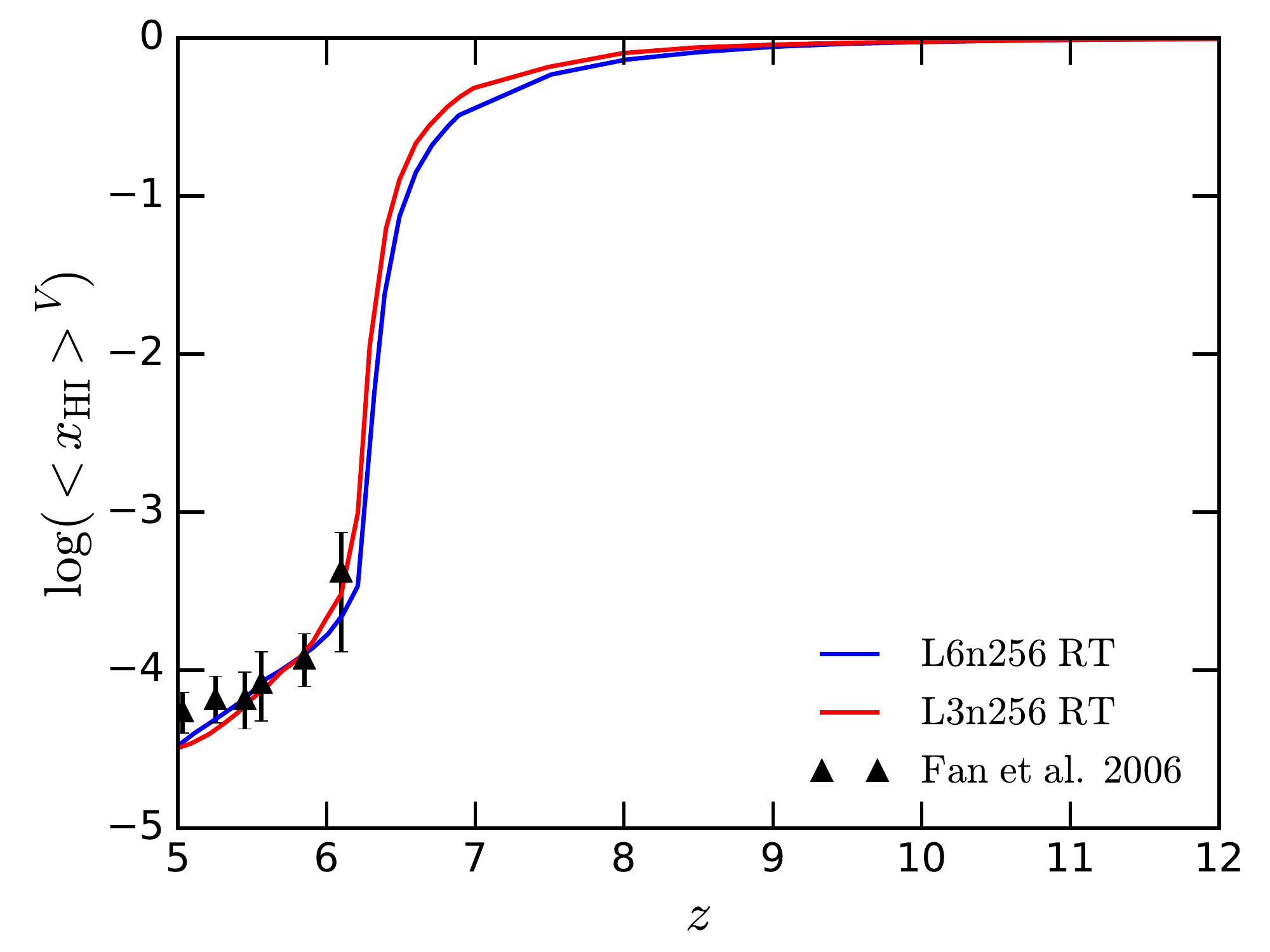}
\caption{Volume-averaged \hi fraction as a function of redshift. Blue and red lines come from L6n256 fiducial-RT and L3n256 fiducial-RT simulations respectively.}
\label{fig:convergence_HI-z}
\end{figure}

We generated three different initial conditions (ICs) for the L3n256 runs using three different random number seeds. An escape fraction of 0.7 is adopted for running all these ICs. Due to the larger cosmic variance of a 3 cMpc$/h$ box, only one of the ICs is able to generate a hydrogen reionization history that experiences complete overlap of ionized bubbles at $z\approx6$. Fig.~\ref{fig:convergence_HI-z} shows the resulting volume-averaged hydrogen ionization fraction as a function of redshift of this IC (red line) compared with that of the L6n256 fiducial-RT run (blue). We will focus the analysis of the remaining of this section on results from this particular IC. The other two ICs give reionization histories that end at $z<5$, because in general L3n256 requires a higher escape fraction to finish reionization at the same redshift as L6n256, due to the lack of bright sources responsible for completing reionization \citep{Katz18} and the ability to resolve more small scale clumping.

\begin{figure*}
\includegraphics[width=2\columnwidth]{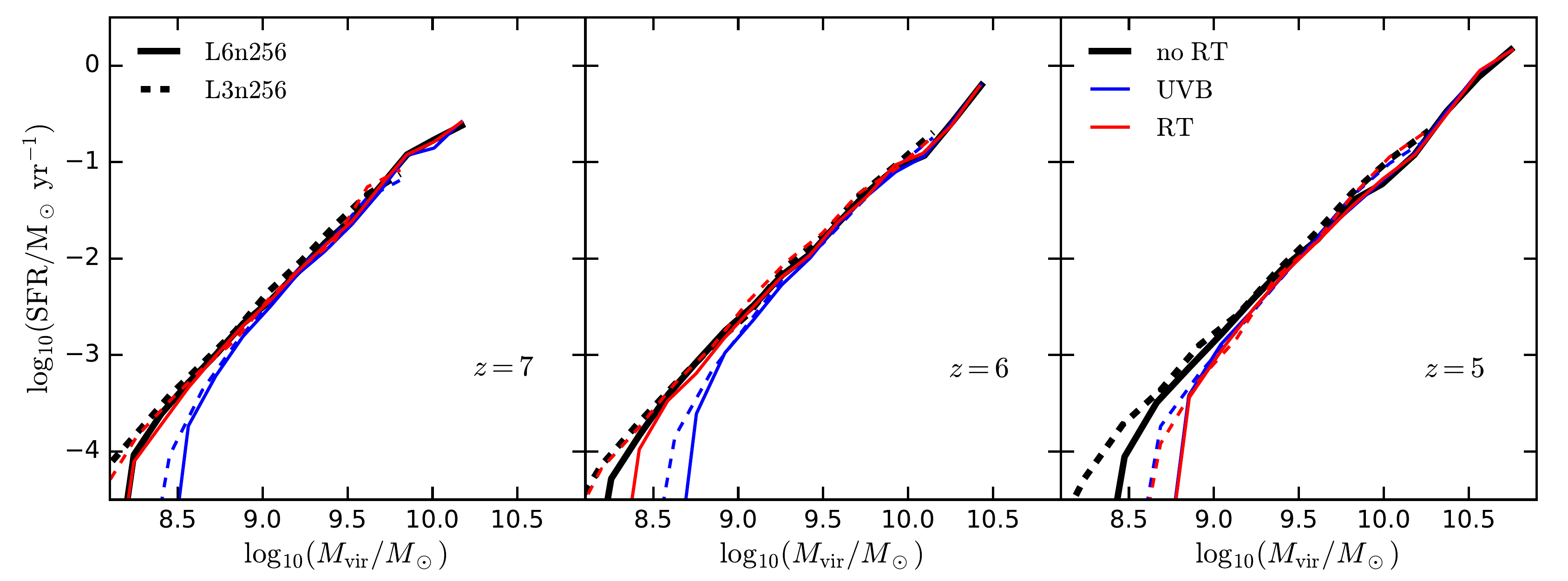}
\caption{Median instantaneous star formation rate versus halo mass relations at $z=7,6,5$. Solid and dashed lines come from L6n256 and L3n256 fiducial runs respectively. Black, blue, and red curves represent the no RT, with FG09 UVB, and with RT variations, respectively. The low-mass halos generally have higher SFR in the L3n256 runs because higher mass resolution resolves more star formation, but the positions of 50\% suppression of SFR due to photoheating feedback are still relatively robust.}
\label{fig:convergence_SFR-Mhalo}
\end{figure*}

Fig.~\ref{fig:convergence_SFR-Mhalo} shows the median SFR as a function of halo mass extracted from L3n256 (dashed lines) and L6n256 (solid lines) at $z=7,6,5$ (from left to right panels). Black, blue, and red lines represent no RT, with FG09 UVB, and with RT respectively. The L6n256-noRT SFR--halo mass relations start turning downwards from the L3n256-noRT ones at $\sim10^{8.2}\ M_\odot$, $\sim10^{8.5}\ M_\odot$, and $\sim10^{8.7}\ M_\odot$ at $z=7,6,5$, respectively. The deviations of the L6n256-UVB run from L3n256-UVB happen at larger halo masses because of the quenching by photoheating. The L3n356-RT run at $z=6$ basically shows no suppression of SFR compared to the L3n256-noRT run, due to the slightly later overlap of ionized bubbles than L6n256-RT. Based on this convergence study, we are more inclined to conclude that at $z=6$ there is little or no suppression of SFR due to photoheating by RT. Nevertheless, at $z=5$ we find good agreement between our L3n256 and L6n256 RT and UVB runs, in terms of the position of 50\% suppression of SFR (halo mass $\sim10^{8.8}\ M_\odot$). Therefore although the low-mass halos may not be completely quenched in the L3n256-RT and L3n256-UVB runs, the suppression of SFR in the L6n256 simulations is indeed a photoheating effect, and it is not caused by lack of resolution.

\begin{figure*}
\includegraphics[width=2\columnwidth]{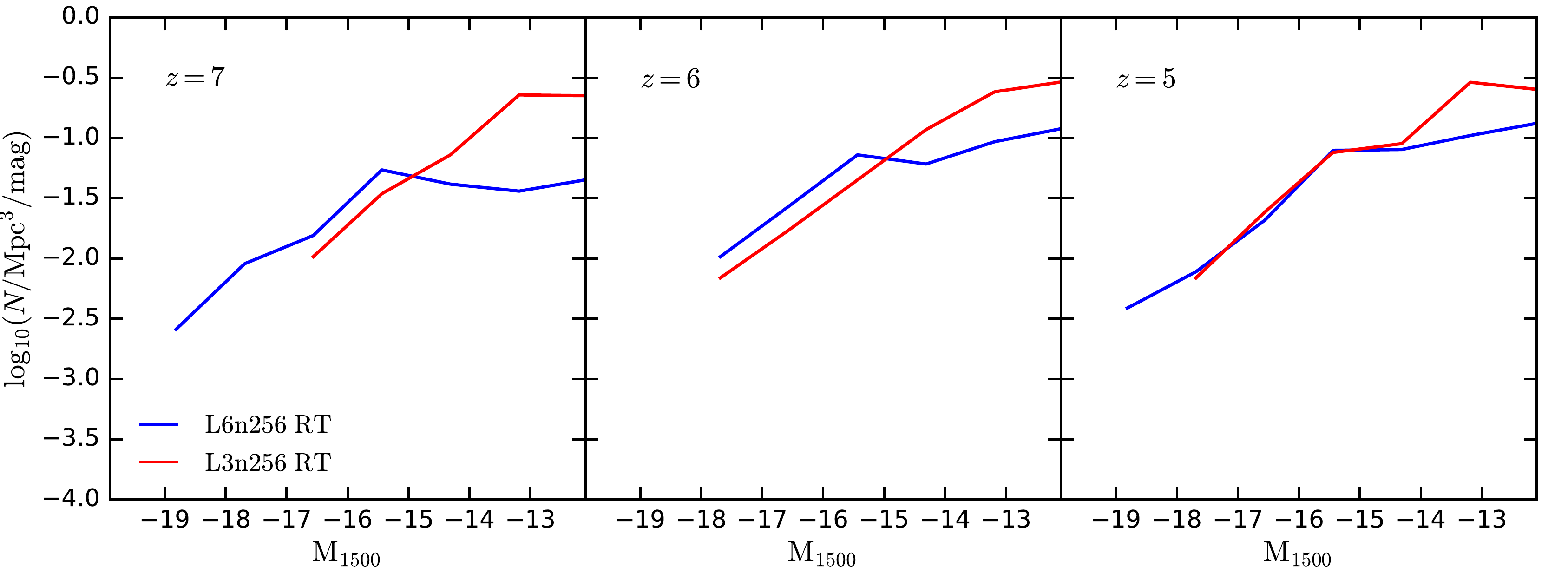}
\caption{UVLFs at $z=7,6,5$, obtained from the L6n256 (blue) and L3n256 (red) fiducial-RT runs. The L6n256 UVLFs turn over at $\sim-15$ mag because of insufficient sampling of the star formation history at a low resolution. The L6n256 UVLFs are therefore relatively robust for magnitudes smaller than $\sim-15$ mag.}
\label{fig:convergence_UVLF}
\end{figure*}

Fig.~\ref{fig:convergence_UVLF} compares the UVLFs from the L6n256-RT (blue) and L3n256-RT (red) runs at $z=7,6,5$ (from left to right). Comparisons between the no RT or UVB runs are similar. Since the L3n256 UVLFs suffer more from stochasticity, we do not try to combine the UVLFs from the two simulation volumes to get a large dynamical range in UV luminosity. At ${\rm {\rm M_{1500}}} \gtrsim -15$ mag, the L6n256 UVLFs experience a turnover compared to the L3n256 ones due to lack of resolution. Increasing the mass resolution by a factor of 8 generates more star particles to sample the star formation history, raising the faint end of the UVLF. We thus cut off the UVLFs at -15 mag in Fig.~\ref{fig:UVLF}.

\section{Numerical convergence of the reduced speed of light approximation}
\label{sec:RSL}

\begin{figure}
\includegraphics[width=\columnwidth]{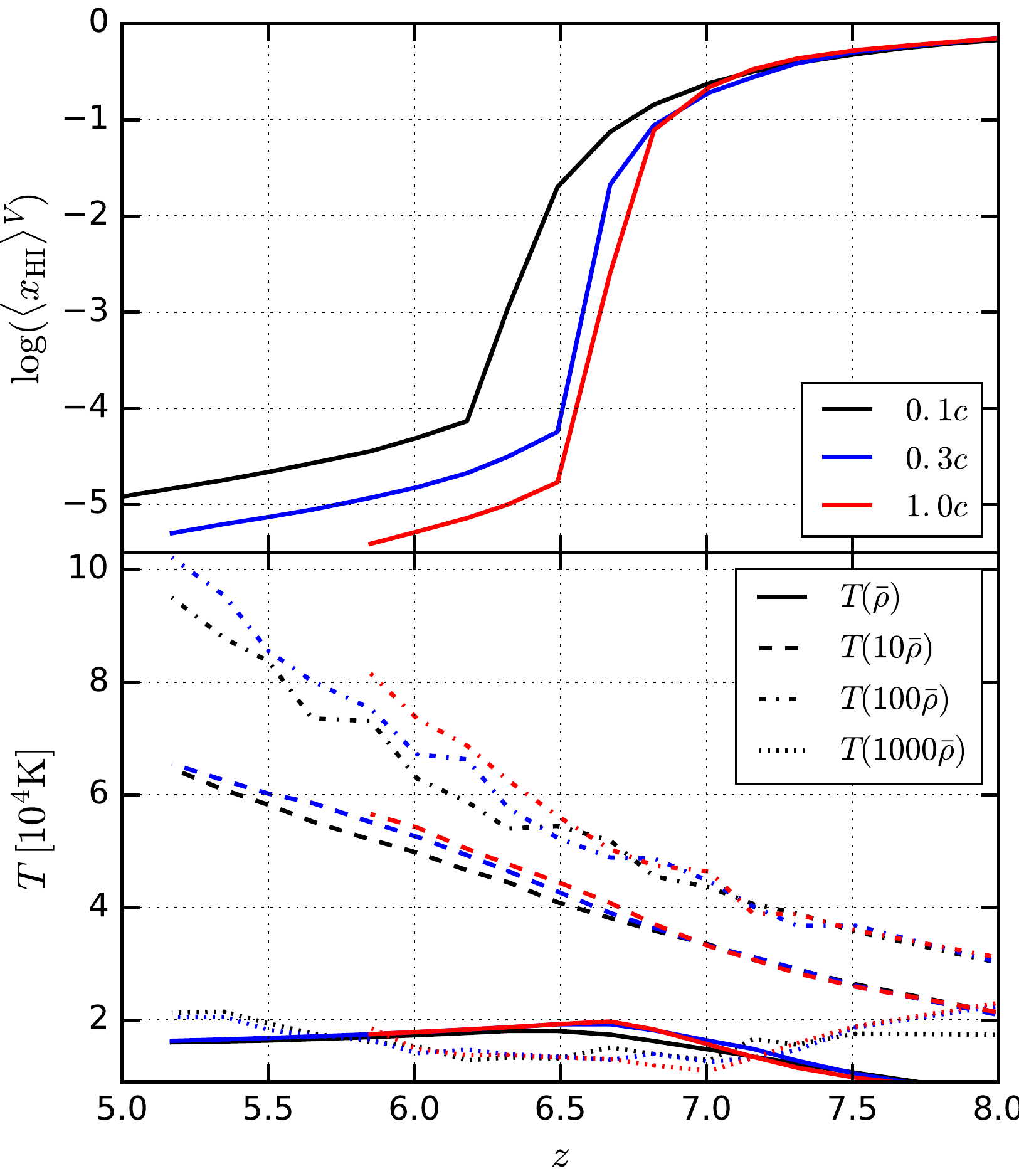}
\caption{The redshift evolution of the volume-averaged \hi fraction (top) and gas temperatures at different overdensities (bottom). Black, blue, and red represent L25n512 RT simulations run with $0.1c$, $0.3c$, $0.5c$, respectively. Solid, dashed, dot-dashed, and dotted lines show gas temperatures at overdensities of $1$, $10$, $100$, $1000$, respectively. Our results are generally consistent with \protect\citet{Depa19} and \protect\citet{Ocvi18RSL}. Using $0.1c$ results in gas temperatures at overdensities of $10-1000$ being underestimated by $5000-10,000$ K after reionization.}
\label{fig:HI-T-z-c}
\end{figure}

To assess the possible effects of using the reduced speed of light on the suppression of halo baryon fraction and SFR, we have performed simulations of $25$ cMpc$/h$ box size with $2\times256^3$ resolution elements (L25n256) with $0.1c$, $0.3c$, and $1.0c$. All three simulations are run with $f_{\rm esc}=1$.
The mass resolution of L25n256 does not allow us to probe the suppression of SFR in $\lesssim10^9\ M_\odot$ halos directly, but these simulations provide information about how the gas temperature changes with the adopted speed of light.
The top panel of Fig.~\ref{fig:HI-T-z-c} shows the resulting $\langle x_{\rm \hi} \rangle^V$ evolution with $z$. Simulations using $0.1c$, $0.3c$, and $1.0c$ are represented by black, blue, and red lines, respectively. The reionization histories are well-converged before $\langle x_{\rm \hi} \rangle^V$ drops to $\sim0.01$. Using $0.3c$ gives good convergence in terms of the time of reaching $\langle x_{\rm \hi} \rangle^V=10^{-4}$, but adopting $0.1c$ delays this redshift by $\sim0.5$. The post-reionization $\langle x_{\rm \hi} \rangle^V$ scales as the inverse of the value of the reduced speed of light. These results are consistent with the findings of \citet{Depa19} and \citet{Ocvi18RSL}.

The bottom panel of Fig.~\ref{fig:HI-T-z-c} illustrates the temperature evolution of gas with overdensities of $1$ (solid), $10$ (dashed), $100$ (dot-dashed), and $1000$ (dotted). The temperature of the IGM does not depend on the amplitude of the post-reionization UVB, so $T(\bar{\rho})$ is well-converged. $T(1000\bar{\rho})$ is also converged at all redshifts, consistent with the findings of \citet{Ocvi18RSL} that $x_{\rm \hi}$ is converged at overdensities $\gtrsim1000$ after reionization. Before $\langle x_{\rm \hi} \rangle^V$ drops to $\sim0.01$, gas temperature at all overdensities are converged. After that point, using $0.1c$ can lead to gas temperatures at overdensities of $10$ and $100$ being underestimated by $5,000-10,000$ K.

\begin{figure}
\includegraphics[width=\columnwidth]{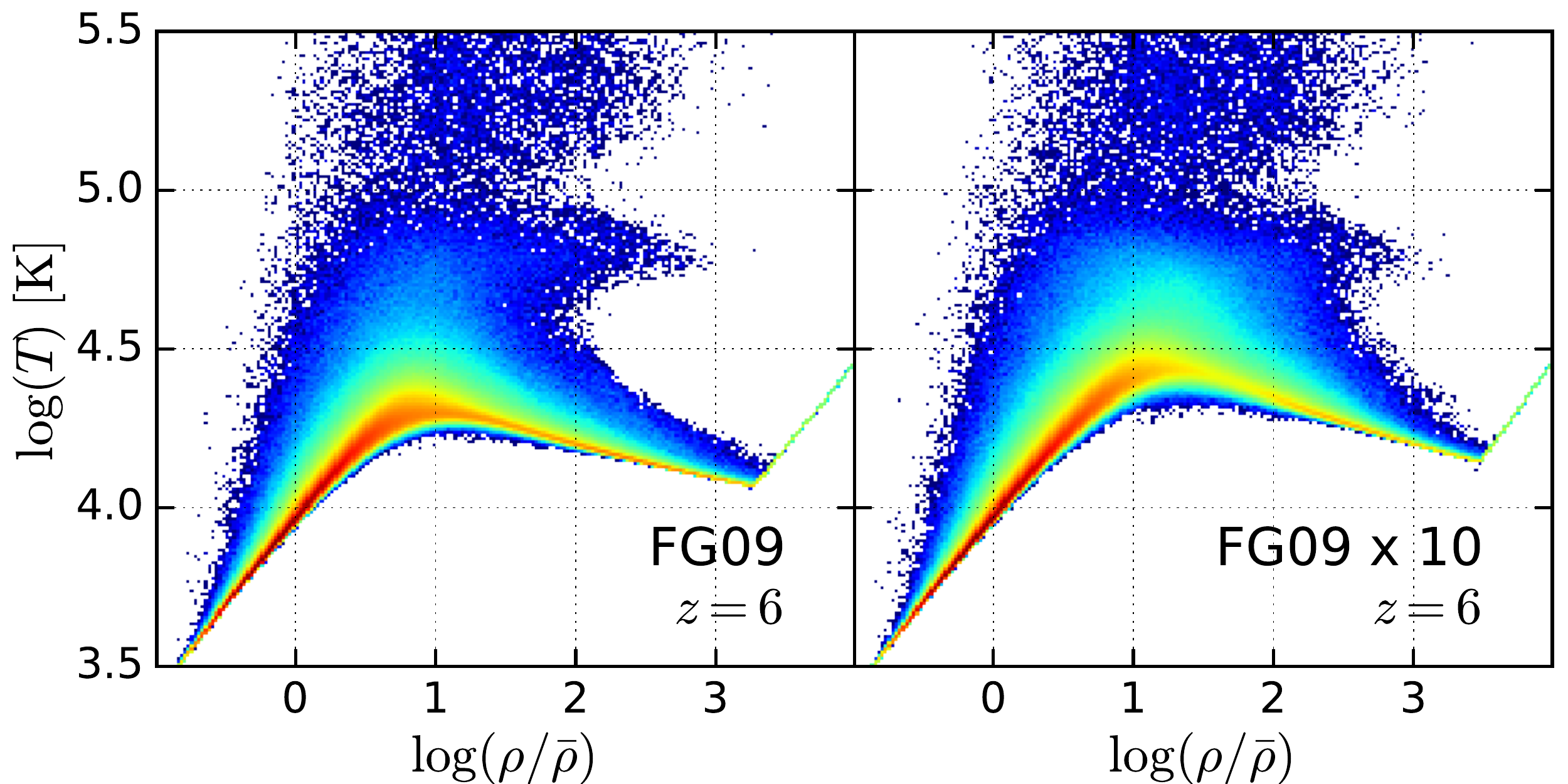}
\caption{Temperature--density diagrams at $z=6$ in the original L6n256 FG09 UVB simulation (left) and the FG09x10 simulation where the photoionization and photoheating rates are scaled by a factor of $10$ (right). Gas with overdensities larger than $\sim10$ have $\sim10,000$ K higher temperatures in FG09x10 than FG09.}
\label{fig:phasediagram_FG09x10}
\end{figure}

\begin{figure}
\includegraphics[width=\columnwidth]{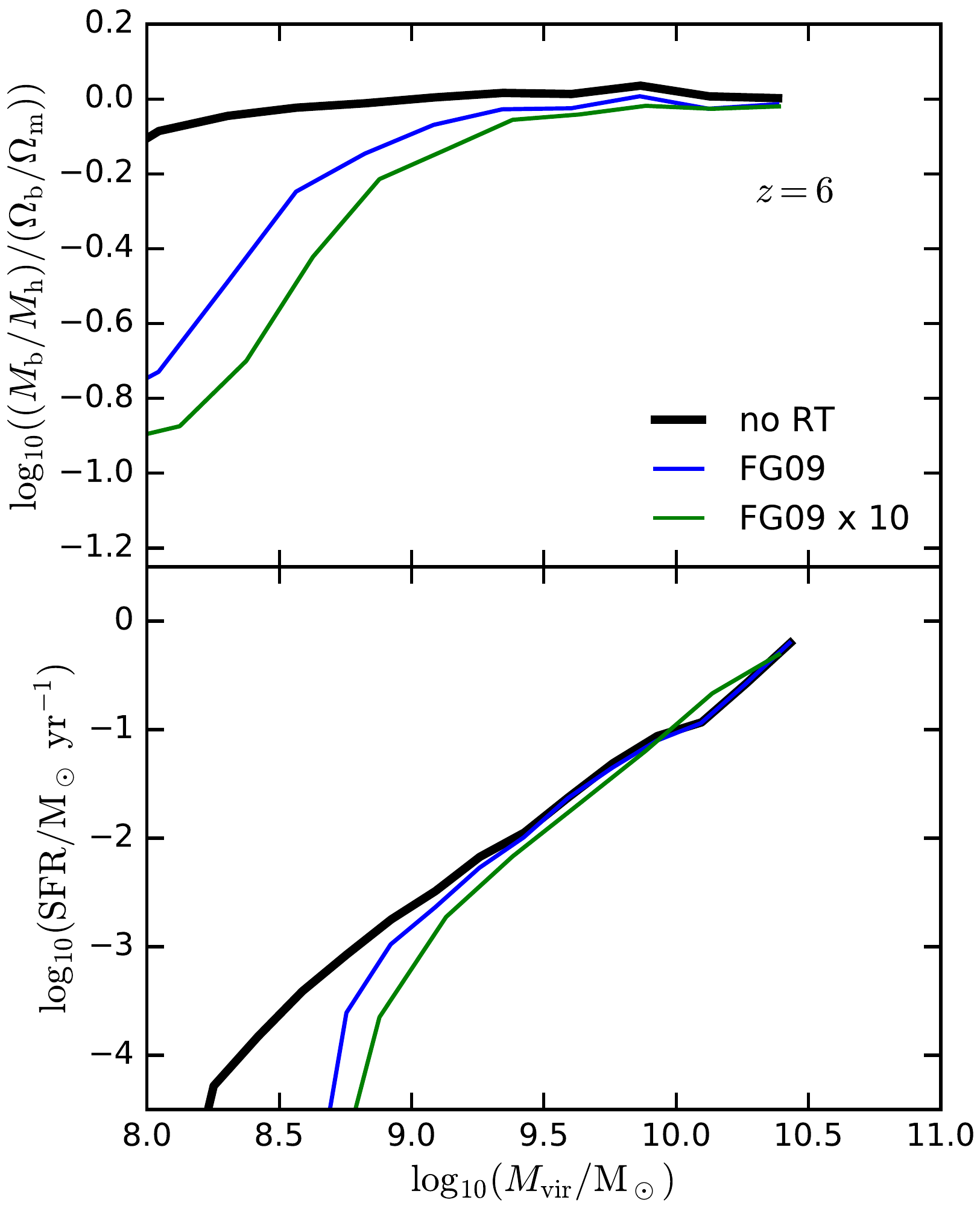}
\caption{Baryon fraction--halo mass relation (top) and SFR--halo mass relation (bottom) at $z=6$ in the no RT (black), FG09 (blue), and FG09x10 (green) simulations. The $\sim10,000$ K temperature difference in the halo gas of FG09 and FG09x10 causes the baryon fraction in low-mass halos to be reduced by $\sim0.2$ dex, and increases the halo mass threshold of the onset of SFR suppression by $0.1\sim0.2$ dex.}
\label{fig:suppression_FG09x10}
\end{figure}

In order to understand how this $5,000-10,000$ K underestimation in the dense gas temperature can affect the suppression of star formation, we ran an additional L6n256 simulation with the FG09 UVB, but with the photoionization and photoheating rates scaled by a factor of $10$ (denoted by FG09x10). This mimics the effects of using the actual speed of light, especially after reionization.
Fig.~\ref{fig:phasediagram_FG09x10} shows the temperature--density diagrams at $z=6$ in the original FG09 simulation (left) and the FG09x10 simulation (right). The IGM temperature is unchanged, as expected. Gas with overdensities larger than $\sim10$ have $\sim10,000$ K higher temperatures in FG09x10 than FG09.
Fig.~\ref{fig:suppression_FG09x10} illustrates the baryon fraction--halo mass relation (top panels) and SFR--halo mass relation (bottom panels) at $z=6$ in the FG09 (blue lines) and FG09x10 (green lines) simulations. The no RT simulation results are shown in black. Despite the $\sim10,000$ K difference in the halo gas temperature, the $10^8\ M_\odot$ halos in FG09x10 only experience a $\sim0.2$ dex more decrease in the baryon content than FG09. The halo mass range of SFR suppression is enlarged by $0.1\sim0.2$ dex in FG09x10, implying the effect of $\sim10,000$ K temperature difference is not strong. We therefore conclude that our results on the suppression of halo SFR is relatively robust with the choice of the reduced speed of light.

\section{IGM Clumping}
\label{sec:clumping}
\begin{figure}
\includegraphics[width=\columnwidth]{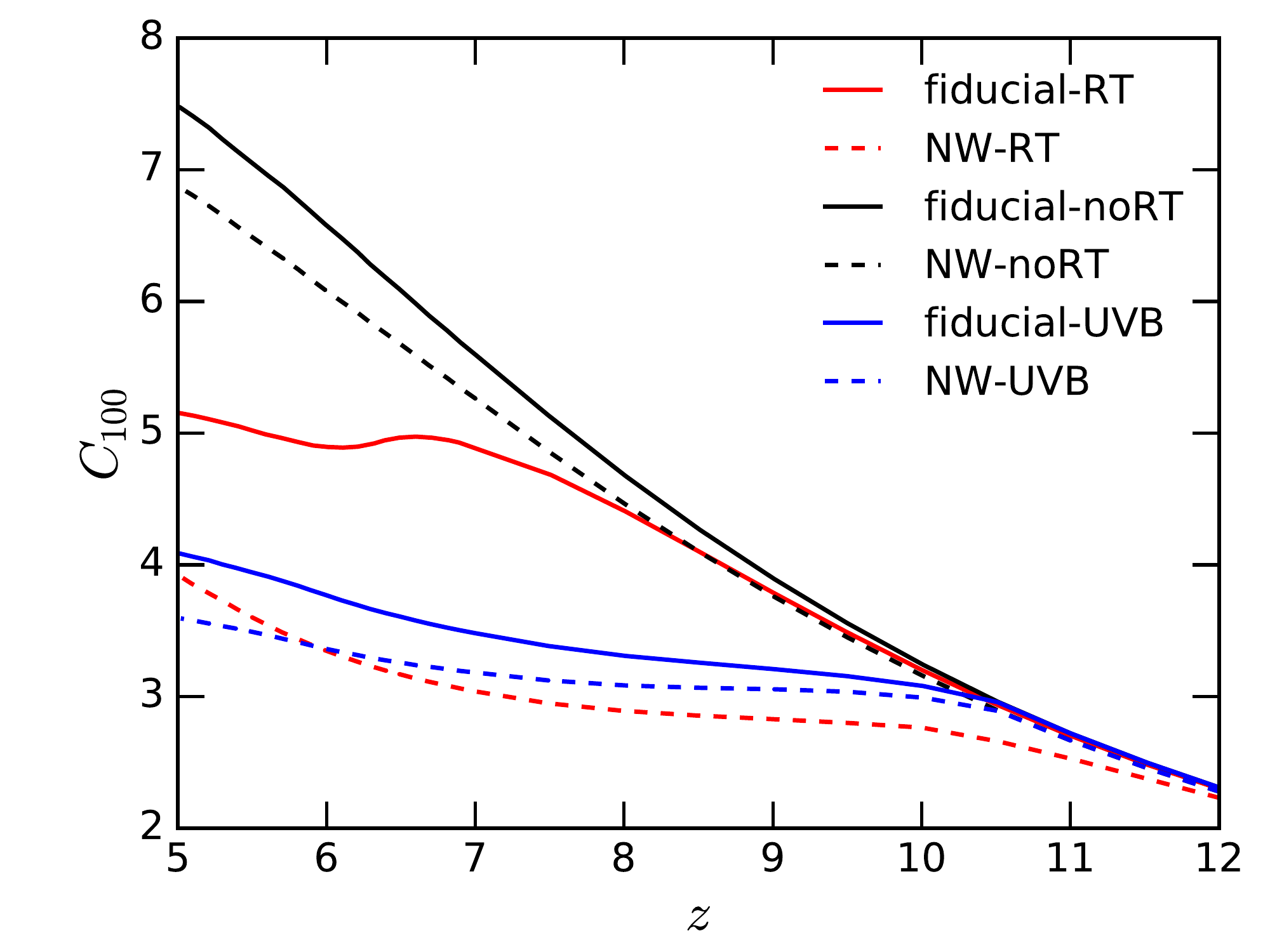}
\caption{Evolution of the IGM clumping factor $C_{100}$, calculated for gas with overdensities $\le100$. Solid and dashed lines represent fiducial and NW runs, respectively. Black, red, and blue show the no RT, RT, and UVB runs, respectively. Photoheating strongly reduces the gas clumping by increasing the Jeans mass, while galactic wind increases $C_{100}$ because it blows dense gas out of galaxies.}
\label{fig:C-z}
\end{figure}

We investigate how photoheating and galactic wind reduce the IGM recombination rate by computing the clumping factor $C = \langle \rho^2 \rangle / \langle \rho \rangle^2$. Here we focus on $C_{100}$, which parametrizes the average recombination rate of gas with overdensities $\le100$. Fig.~\ref{fig:C-z} shows the evolution of $C_{100}$ with redshift in the fiducial (solid lines) and NW (dashed lines) simulations. Black, red, and blue represent no RT, RT, and UVB runs, respectively. Photoheating strongly decreases the clumping factor by increasing the Jeans mass of the ionized gas. Galactic wind raises the clumping factor because it blows gas out of galaxies into the IGM. The fiducial-UVB run also generates much lower $C_{100}$ than the fiducial-RT run because the FG09 UVB turns on at a high redshift. These results are consistent with the findings of \citet{Pawl15}, although the NW-RT run produces a much lower $C_{100}$ than their corresponding simulation because of a much earlier reionization. The effects of photoheating are also qualitatively similar to what was found by \citet{Finl12}, but their galactic wind model does not seem to move gas out of galaxies as efficiently as ours. We therefore conclude that the simulated effects of photoheating on the IGM properties are relatively robust.

\bsp	
\label{lastpage}
\end{document}